\documentclass[conference, compsoc]{IEEEtran}
\newcommand{\cure}{CAT}
\PassOptionsToPackage{usenames,dvipsnames}{xcolor}
\usepackage{hyperref}
\usepackage{array, multirow}
\newcolumntype{P}[1]{>{\centering\arraybackslash}p{#1}}
\usepackage{tikz}
\usepackage{listings}

\definecolor{codebg}{HTML}{F7F7F9}
\definecolor{codeframe}{HTML}{E1E4E8}
\definecolor{codekw}{HTML}{005CC5}      
\definecolor{codectrl}{HTML}{AA0D91}    
\definecolor{codestring}{HTML}{22863A}
\definecolor{codecomment}{HTML}{6A737D}
\definecolor{codenumber}{HTML}{8A8F98}
\lstdefinelanguage{Pseudo}{
  morekeywords=[1]{for,in,do,end,return,while,if,then,else,elseif,break,continue},
  morekeywords=[2]{function,procedure,proc,def,var,const,let},
  morekeywords=[3]{true,false,null,None},
  sensitive=false,
  morecomment=[l]{//},
  morecomment=[s]{/*}{*/},
  morestring=[b]"
}

\lstdefinestyle{clean}{
  language=Pseudo,
  literate=
    *{0}{{\textcolor{darkred}{0}}}1
    {1}{{\textcolor{darkred}{1}}}1
    {2}{{\textcolor{darkred}{2}}}1
    {3}{{\textcolor{darkred}{3}}}1
    {4}{{\textcolor{darkred}{4}}}1
    {5}{{\textcolor{darkred}{5}}}1
    {6}{{\textcolor{darkred}{6}}}1
    {7}{{\textcolor{darkred}{7}}}1
    {8}{{\textcolor{darkred}{8}}}1
    {9}{{\textcolor{darkred}{9}}}1,
  basicstyle=\fontsize{8}{5}\selectfont\ttfamily,
  rulecolor=\color{codeframe},
  numbers=left,
  numberstyle=\scriptsize\color{codenumber},
  numbersep=8pt,
  showstringspaces=false,
  columns=fullflexible,
  keepspaces=true,
  tabsize=2,
  breaklines=true,
  aboveskip=0.8\baselineskip,
  belowskip=0.8\baselineskip,
  keywordstyle=[1]\bfseries\color{codekw},
  keywordstyle=[2]\bfseries\color{codectrl},
  keywordstyle=[3]\bfseries\color{Teal},
  commentstyle=\itshape\color{codecomment},
  stringstyle=\color{codestring}
}
\definecolor{codegray}{rgb}{0.1,0.1,0.1}

\definecolor{darkblue}{RGB}{20, 20, 139}
\definecolor{darkgreen}{RGB}{0, 100, 0}
\definecolor{darkred}{RGB}{139, 0, 0}
\definecolor{darkorange}{RGB}{215, 110, 0}

\definecolor{darkpurple}{RGB}{75, 0, 130}
\definecolor{darkgray}{RGB}{105, 105, 105}
\definecolor{darkcyan}{RGB}{0, 139, 139}
\definecolor{darkteal}{RGB}{0, 128, 128}

\lstdefinestyle{compact}{
    keywordstyle=\color{darkblue},
    commentstyle=\color{darkgreen},
    stringstyle=\color{darkorange},
    literate=
        *{0}{{\textcolor{darkred}{0}}}1
        {1}{{\textcolor{darkred}{1}}}1
        {2}{{\textcolor{darkred}{2}}}1
        {3}{{\textcolor{darkred}{3}}}1
        {4}{{\textcolor{darkred}{4}}}1
        {5}{{\textcolor{darkred}{5}}}1
        {6}{{\textcolor{darkred}{6}}}1
        {7}{{\textcolor{darkred}{7}}}1
        {8}{{\textcolor{darkred}{8}}}1
        {9}{{\textcolor{darkred}{9}}}1,
    numberstyle=\tiny\color{codegray},
    basicstyle=\fontsize{8}{5}\selectfont\ttfamily,
    columns=fullflexible,
    breakatwhitespace=false,
    breaklines=true,                 
    captionpos=b,                    
    keepspaces=true,                 
    numbers=left,                    
    numbersep=3pt,                  
    showspaces=false,                
    showstringspaces=false,
    showtabs=false,                  
    tabsize=2,
    frame=none, 
    framexbottommargin=0pt, 
    framextopmargin=0pt, 
}

\lstdefinestyle{compact-alt}{
    keywordstyle=\color{darkpurple},
    commentstyle=\color{darkgray},
    stringstyle=\color{darkteal},
    literate=
        *{0}{{\textcolor{darkcyan}{0}}}1
        {1}{{\textcolor{darkcyan}{1}}}1
        {2}{{\textcolor{darkcyan}{2}}}1
        {3}{{\textcolor{darkcyan}{3}}}1
        {4}{{\textcolor{darkcyan}{4}}}1
        {5}{{\textcolor{darkcyan}{5}}}1
        {6}{{\textcolor{darkcyan}{6}}}1
        {7}{{\textcolor{darkcyan}{7}}}1
        {8}{{\textcolor{darkcyan}{8}}}1
        {9}{{\textcolor{darkcyan}{9}}}1,
    numberstyle=\tiny\color{codegray},
    basicstyle=\fontsize{7.3}{5}\selectfont\ttfamily,
    columns=fullflexible,
    breakatwhitespace=false,
    breaklines=true,                 
    captionpos=b,                    
    keepspaces=true,                 
    numbers=left,                    
    numbersep=3pt,                  
    showspaces=false,                
    showstringspaces=false,
    showtabs=false,                  
    tabsize=2,
    frame=none, 
    framexbottommargin=0pt, 
    framextopmargin=0pt, 
}

\usepackage{makecell}
\usepackage{subcaption}
\usepackage{flushend}
\usepackage{fancyhdr}
\usepackage{algorithm2e}
\usepackage{cprotect}
\usepackage{environ}
\usepackage{svg}
\usepackage{xcolor}
\usepackage{graphicx}

\newcommand{\updated}[1]{\color{black}#1}

\newcolumntype{L}[1]{>{\raggedright\let\newline\\\arraybackslash\hspace{0pt}}m{#1}}
\newcolumntype{C}[1]{>{\centering\let\newline\\\arraybackslash\hspace{0pt}}m{#1}}
\newcolumntype{R}[1]{>{\raggedleft\let\newline\\\arraybackslash\hspace{0pt}}m{#1}}
\usepackage{algorithmicx}

\usepackage{graphicx}
\usepackage{caption}
\renewcommand{\arraystretch}{1.25}
\usepackage{pifont}

\newcommand{\ignore}[1]{}

\newcommand{\nv}[1]{\textcolor{black}{#1}}

\newcolumntype{P}[1]{>{\centering\arraybackslash}p{#1}}
\usepackage{tikz}

\newcolumntype{H}{>{\setbox0=\hbox\bgroup}c<{\egroup}@{\hspace*{-\tabcolsep}}}

\newcolumntype{C}[1]{>{\centering\arraybackslash}p{#1}}

\begin{document}

\author{
\IEEEauthorblockN{Haya Schulmann and Niklas Vogel}
\IEEEauthorblockA{Goethe-Universität Frankfurt, Germany \\ ATHENE National Research Center for Applied Cybersecurity, Germany}
}

\fancyhf{} 
\fancyfoot[C]{\thepage} 
\title{Batch Me If You Can: Coverage-guided RPKI Fuzzing at Scale}
\maketitle
\begin{abstract}
The Resource Public Key Infrastructure (RPKI) has become essential to secure inter-domain routing. Despite its critical role, RPKI software remains largely untested beyond shallow parsing. Existing fuzzers, like AFL++ or libFuzzer, do not work well for RPKI as they assume a single, self-contained input per execution, while RPKI repositories contain hundreds of interdependent cryptographically linked objects. 
Existing fuzzers fail to handle this complexity and lack the ability for precise coverage attribution in multi-object repositories, breaking feedback-based exploration and thereby missing most severe vulnerabilities in RPKI validation.\\
\indent In this paper, we overcome these limitations through novel fuzzing techniques, 
including continuous sampling and using functions as side-channels for per-object coverage attribution in large input repositories. We further show how parsing inputs to a labeled tree allows structural and semantic mutations while preserving cryptographic validity in mutated repositories.
We implement our new techniques into a powerful fuzzing tool called \cure, combining non-sequential fuzzing with our template-agnostic ASN.1 mutation engine to achieve 66× throughput improvement over sequential fuzzing and exploring 24 -- 47\% more unique code paths compared to libFuzzer and previous work.\\
\indent Evaluating \cure\ on RPKI validators 
uncovered 21 previously unknown vulnerabilities with 8 CVEs already assigned  (CVSS 7.5 -- 9.8). These include a buffer overflow, Denial-of-Service (DoS), and exploitable repository-poisoning logic flaws.\\
\indent We open-source \cure\ to enable reproducibility, further research, and adaptation of our methods to other complex cryptography-based protocols such as DNSSEC and TLS.
\end{abstract}

\section{Introduction}
The Resource Public Key Infrastructure is now a cornerstone of secure inter-domain routing, protecting the Border Gateway Protocol (BGP) against prefix hijacks \cite{DBLP:journals/cacm/SunABVRCM21}. With U.S. federal agencies and the White House urging broader adoption \cite{FCC2024bgp,ONCD2024BGPRoadmap}, RPKI validators are becoming critical infrastructure and potential high-value attack targets. Yet, despite their importance, their robustness remains largely untested beyond shallow parsing checks.  
Fuzzing has proven effective for uncovering flaws in complex protocols, but RPKI presents unique obstacles: validation depends on strongly interlinked ASN.1 objects, each with cross-referenced hashes, serials, and signatures. This cryptographic verification imposes high per-object setup costs, and multi-object dependencies frustrate the sequential, single-input model of popular coverage-guided fuzzers like AFL++. As a result, existing methods either fail early in parsing and validation or lose the coverage feedback needed to reach deep validation logic, leaving critical bugs undiscovered. At the same time, the complexity of RPKI also makes correct implementation of RPKI challenging and error-prone, making sophisticated testing essential to discover critical flaws. 
As we show in this work, weaknesses and bugs in RPKI can degrade the protection of otherwise secure systems and even expose the servers running RPKI software to direct attacks. Securing BGP through RPKI ultimately depends on the robustness of RPKI itself.

\textbf{Challenges of fuzzing complex cryptographic architectures.}
RPKI fuzzing is challenging due to its intricate interdependent validation routines, secure key management, parsing large cryptographic objects, and establishing trust chains across multiple entities. These challenges extend beyond RPKI, which is why most protocol fuzzing to date has focused on non-cryptographic ``vanilla'' versions, like DNS or BGP, with lower complexity \cite{zhang2023resolverfuzz,bh:fuzz:bgp,singha2024messi}. Even recent fuzzing work on DNS explicitly disabled DNSSEC due to its cryptographic complexity \cite{zhang2023resolverfuzz}. In the 26 years since the introduction of DNSSEC (RFC2535), no fuzzing studies have targeted it. Similarly, the first attempt to fuzz RPKI appeared only recently \cite{mirdita2023cure} and as we explain in \S\ref{sc:works}, it relied on rudimentary techniques, overlooking severe vulnerabilities which we find in this work. The lack of rigorous testing for these protocols is concerning, as critical flaws can remain undiscovered for decades \cite{heftrig2024harder}.\\
\indent A fundamental limitation of existing tools is that they are exclusively sequential: tools like AFL++ or libFuzzer test one input, evaluate its performance, and then select/mutate/evaluate the next input. Sequential operation is fundamentally built into these tools since they rely on coverage-guidance of individual objects to score performance. After executing the target with one input, the fuzzer checks explored paths and rates performance accordingly. This is not possible if multiple inputs are tested in parallel, as final coverage performance would be the union of each input, making rating the individual inputs impossible.\\
\indent {\updated The limitation of existing tools requiring sequential operation becomes prohibitive when complex multi-input setups are intrinsic to core protocol validation logic, as in RPKI. Validation of a single RPKI object requires cross-validation with up to 7 additional objects and object validation incurs a substantial setup overhead. This form of cryptographic cross-object validation is a recurring pattern in cryptography-heavy systems: TLS certificate chains incur multiple cross-checking validation steps, validating child certificates against respective parents. DNSSEC chain validation includes DNSKEY hash checks from Delegated Signer (DS) records in the parent, and intra-zone signature validations depend on the key record, signature record, and the respective signed record set. 
Re-generating and processing the setup for each input makes fuzzing slow and inefficient, and fails to capture real-world multi-object setups effectively.}\\
\indent As we show in this work, batching many inputs in one test-run improves explored paths and decreases setup overhead per input (66x in RPKI), making fuzzing substantially more powerful, but {\updated precludes} sequential fuzzing, making existing tooling inapplicable.\\
\indent In this work, we overcome the limitations of sequential fuzzing by developing a non-sequential fuzzer with coverage-guidance. For this, we make several conceptual and technical contributions:\\
\indent \textbullet\ We develop new techniques for object-specific coverage attribution in large input batches with an accuracy of over 99\%. {\updated This contribution enables the first effective coverage-guided fuzzing of RPKI, substantially improving over previous RPKI fuzzing work \cite{mirdita2023cure}.} No prior work has combined high-throughput batching with precise per-input coverage attribution, making this a key novelty and an enabling technique for effective fuzzing of RPKI and other multi-object, cryptography-based protocols.

\textbullet\ We implement and open-source the first template agnostic ASN.1 parsing and mutation module. The module parses any ASN.1 structure without templates, dynamically labels every field path, and applies both structural and semantic mutations while automatically repairing dependent fields and re-signing objects to preserve cryptographic validity across the repository. Mutated ASN.1 objects are encoded with our custom DER encoder to support encoding objects with broken structure.  
Existing fuzzers typically rely on general-purpose encoders and hard coded message templates bound to a single schema, which limits coverage to predefined fields, cannot handle cross-type mutations, and requires manual updates for protocol changes. 
In contrast, our schema-free ASN.1 mutation capability that supports high-throughput allows us generation of valid, signed, and interlinked RPKI repositories and is thus essential for reaching deep validation logic in RPKI fuzzing.

\textbullet\ We develop a novel fuzzing tool, \cure, that incorporates our new techniques to enable efficient coverage-guided RPKI fuzzing. CAT follows a modular design, with general purpose modules for ASN.1 parsing, mutation, and coverage scoring, allowing extension of these modules to other protocols. We open-source \cure\ to facilitate reproducibility of our results and future research.\footnote{\href{https://github.com/niklbird/cat_fuzzer}{https://github.com/niklbird/cat\_fuzzer}}  

\textbullet\ We evaluate \cure\ on all five major RPKI software packages. Using \cure, we run 300 million tests and identify 21 previously unknown vulnerabilities across all tested clients, which we responsibly disclose. Our fuzzing techniques allow us to discover not only DoS vulnerabilities, which are frequently found with fuzzers, but {\updated severe memory-safety} vulnerabilities, such as a buffer overflow we exploit for Remote Code Execution (RCE) to take over the machine running RPKI. Additionally, we find remote cache poisoning attacks that allow stealthy targeted attacks on arbitrary certificate authorities and are much more difficult to detect than crashes of RPKI. Our findings present the first attacks on RPKI that expose to malicious control over the attacked server.

\textbullet\ We provide concrete directions for future research to utilize our techniques for advanced fuzzing of other cryptography-based protocols to improve their security. 

\indent {\bf Organization.} We review RPKI in \S\ref{sc:overview}, compare our research to related work in \S\ref{sc:works} and discuss limitations of existing tools in \S\ref{sec:challenges}. 
We present our non-sequential techniques in \S\ref{sc:non_seq}, discuss the design of \cure\ in \S\ref{sc:architecture}, evaluate \cure\ on RPKI implementations in \S\ref{sc:eval}, and report the vulnerabilities we found with \cure\ in \S\ref{taxonomy}. We provide future research directions in \S\ref{sc:future} and conclude in \S\ref{sc:conclusions}.

\section{Overview of RPKI}\label{sc:overview}
\begin{figure}[t]
    \centering
    \includegraphics[width=1.0\columnwidth]{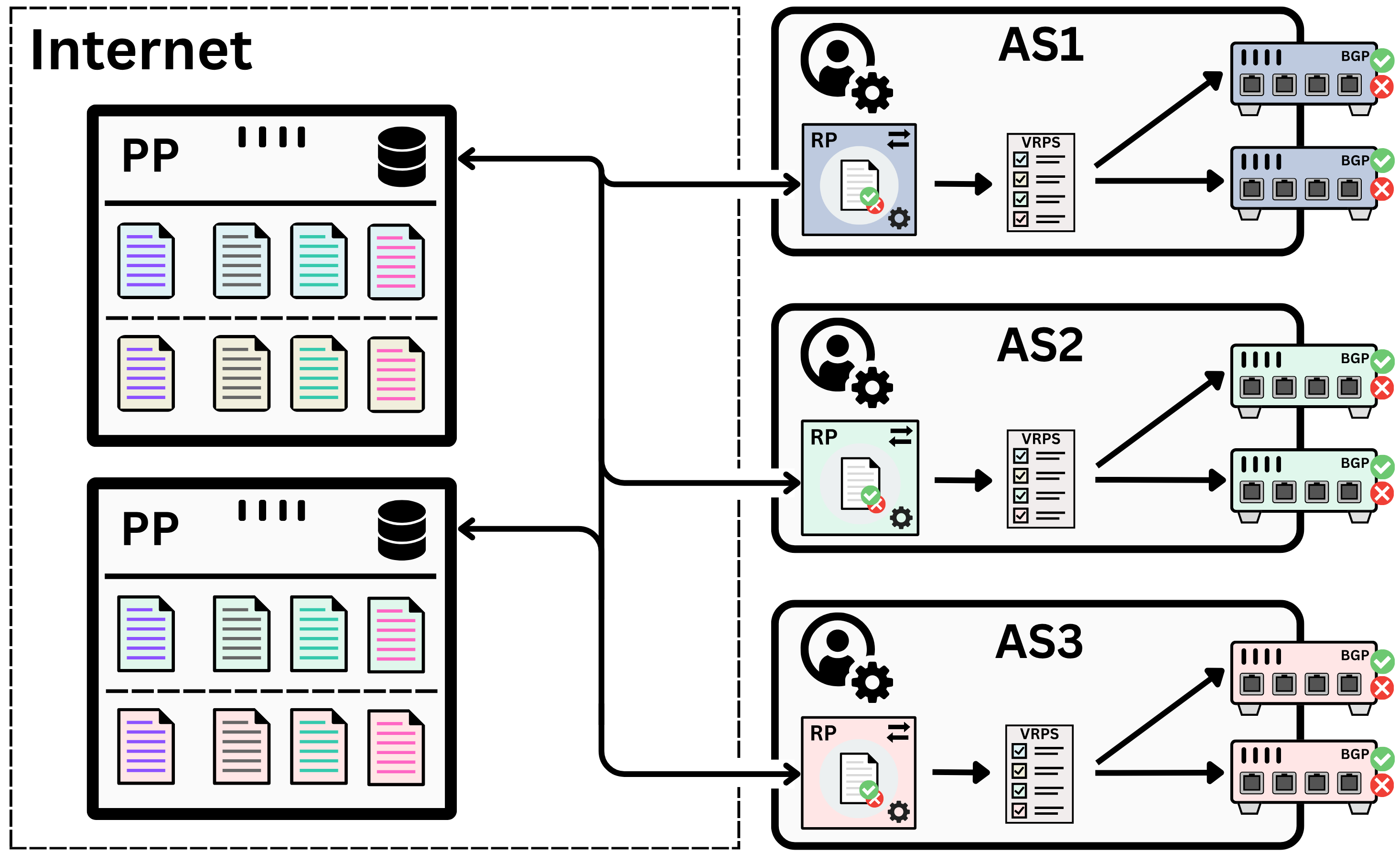}
    \caption{Overview RPKI.}
    \label{fig:rpki_overview}
\end{figure}

RPKI (RFC6480) standardizes an infrastructure to distribute authenticated IP prefix ownership information globally, allowing routers to make secure routing decisions in BGP. An overview of RPKI is given in Figure~\ref{fig:rpki_overview}.
In RPKI, hierarchically organized Publication Points (PPs) store cryptographic RPKI objects. Most importantly, PPs store signed Route Origin Authorizations (ROAs) that detail which Autonomous System (AS) is authorized to announce specific prefixes in BGP (RFC4271). Additional files in the RPKI include certificates, Certificate Revocation Lists (CRLs), Manifests (MFTs), Autonomous System Provider Authorizations (ASPAs), BGPsec certificates, and GhostBuster Records (GBRs).

Routers do not directly interact with RPKI repositories to download/validate these objects. Instead, downloading and validation of objects is performed by Relying Party (RP) validator implementations (RFC6481). 
The workflow of RPs for downloading and validation includes multiple steps: recursive data fetching from PPs 
over the RPKI Repository Delta Protocol (RRDP), parsing and validating RPKI objects, and compiling the validated content for connected routers to make secure routing decisions in BGP. 

\textbf{Complex object validation.}
All objects in RPKI are signed by the private key of the Certificate Authority (CA) managing the resources detailed in the object. RPs perform complex validation steps to validate the RPKI objects. For each CA, the RP constructs and validates a certificate path to the trust root. Each CA links to one signed CRL and one signed manifest, both of which are validated through the CA key using validation steps outlined in RFC6486/RFC6487. The manifest contains a full signed X.509 certificate for validation, and must contain the hashes of all objects issued by the CA. The RP must further confirm which certificates are revoked through the CRL. 

\begin{figure}[t!]
    \centering
    \includegraphics[width=1.0\columnwidth]{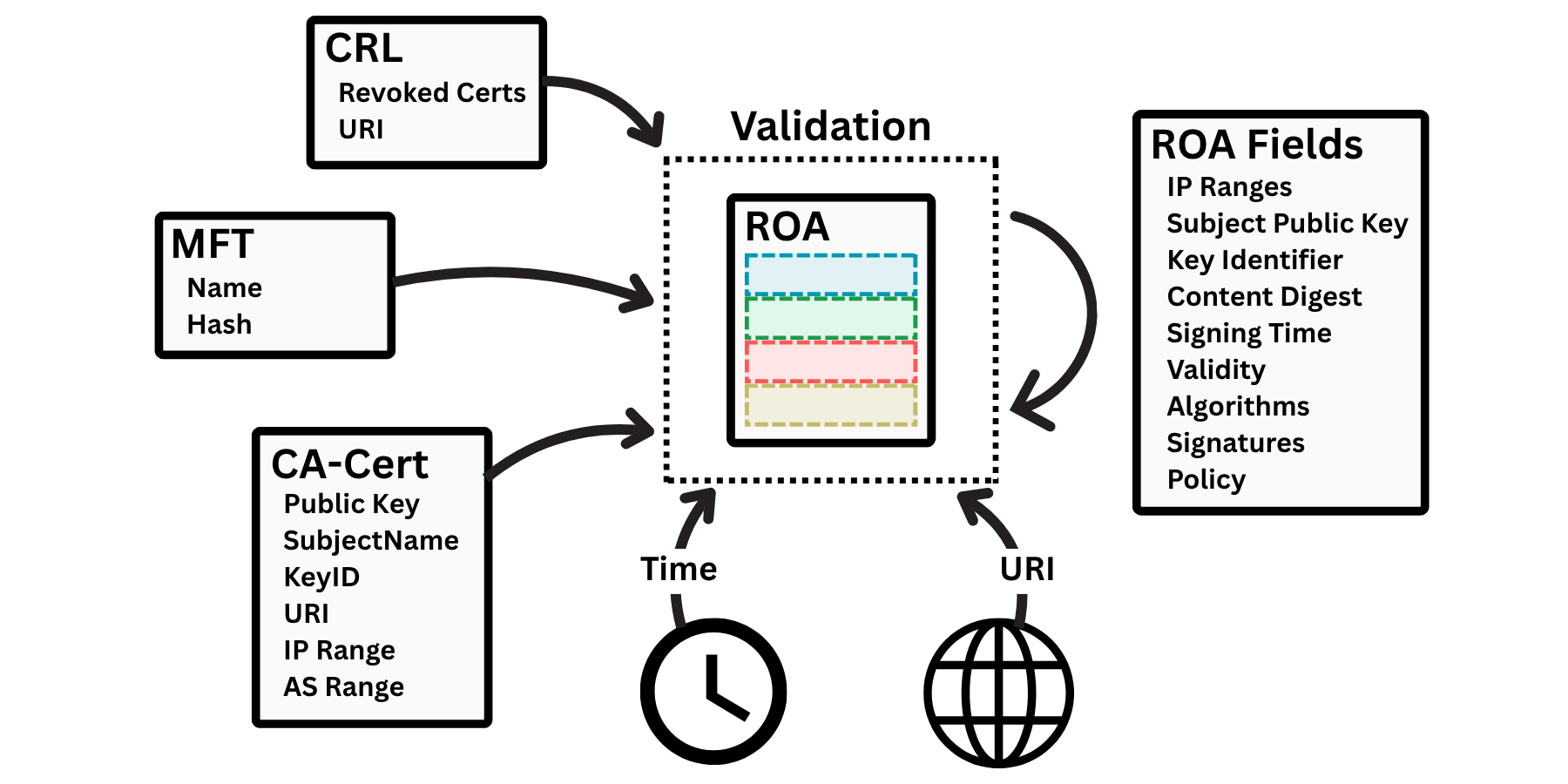}
    \caption{Required fields for ROA validation.}
    \label{fig:roa_validation}
\end{figure}

All other signed objects, like ROAs, contain two signatures, the object’s payload is signed with a one-off key pair, and the one-off public key is contained in the ROA file inside an X.509 certificate. This End-Entity (EE) certificate within the ROA is signed by the CA, so the RP first validates the payload signature with the one-off public key, then validates the EE-certificate's signature with the CA certificate. The RP must additionally validate a substantial amount of fields in the EE certificate, like issuer name, parent fingerprint, or object URI, outlined in RFC6480. The required fields and objects for validation of a single ROA are illustrated in Figure~\ref{fig:roa_validation}. Validation of each ROA thus not only requires two signature validations, but also validation using multiple fields sourced from other RPKI objects by the same CA. This interdependency means that validation of any RPKI object always requires a complex multi-object cryptographic setup including a CA certificate, CRL, and manifest.

\section{\hspace{1mm}Related Work}\label{sc:works}
Fuzzing has been applied to various Internet protocols, including DNS, HTTP, RPKI and TLS \cite{wang2013rpfuzzer,beurdouche2015,liu2020,kakarla2022scale,bushart2023resolfuzz}. 
Each protocol presents unique challenges and obstacles, and the effectiveness of fuzzing techniques varies accordingly. We compare related work with our research, highlighting the key differences and challenges, and explaining how our research resolves the existing limitations.

{\bf Fuzzing techniques.} Fuzzing approaches differ mainly in how they generate inputs and obtain feedback. In black-box fuzzing
inputs are generated without knowledge of the targets internals and without instrumentation. It offers simplicity but no guidance toward unexplored code paths \cite{takanen2018,guo2024stateful}. Grey-box fuzzing instruments the target to collect lightweight feedback such as code coverage and uses it to guide mutations \cite{godefroid2012}. Coverage-guided fuzzing (CGF) is a common form implemented by tools such as AFL and libFuzzer \cite{AFLplusplus-Woot20} to improve fuzzing success. We build upon the approach used by these fuzzers to extract coverage information from a binary. Our work builds on CGF because it provides execution feedback that can steer exploration into deep validation logic, essential for RPKI, where most critical bugs occur beyond initial parsing. Traditional CGF tools assume a single, self-contained input per execution, making them unsuitable for RPKI’s multi-object, cryptographically linked input model. Black-box fuzzing lacks coverage feedback, leading to shallow exploration, while grammar/template-based fuzzing can generate valid ASN.1 but cannot prioritize mutations that reach new code paths. \cure\ retains CGF's feedback-based search but extends it with batching to amortize expensive repository initialization across many objects, and per-object coverage attribution to maintain guidance in a multi-object setting. These extensions address the key limitations that prevent existing fuzzers from effectively testing RPKI validators.

{\bf Previous work on fuzzing network security protocols.} Multiple previous works discuss fuzzing network security protocols, with a major research focus on TLS fuzzing \cite{zhu2020guided, ammann2024dy, chen2015guided, brubaker2014using}. TLS fuzzing is closely related to RPKI, as both use ASN.1 objects and X.509 certificate structure. 

Mucert \cite{chen2015guided} introduced a methodology to fuzz TLS certificates through a corpus of existing certificates mutated with Markov Chain Monte Carlo sampling, improving over the popular Frankencerts approach \cite{brubaker2014using}. However, their fuzzing is limited since their mutation algorithm requires \emph{valid} certificates, omitting the ability to discover complex vulnerabilities that require repeated structure-breaking mutations. Further, their approach is slow, with only 5000 mutated certificates tested. Our work introduces the new technique of labeled ASN.1 syntax trees to support repeated mutations on ill-structured objects. Further, our approach is fast, allowing us to test millions of objects. 

Recent work \cite{ammann2024dy} has deployed Dolev-Yao (DY) models to fuzz TLS implementations and discovered four novel vulnerabilities. However, their approach is targeted at fuzzing the message exchange of TLS server / client, with only limited DY-level mutation capabilities that keep message structure intact. Further, their work requires substantial manual effort to adapt their tool to new targets. The work also does not include new techniques to overcome aspects of TLS with large per-input overhead, like certificate chains. In contrast, our work supports a wide-range of structure-preserving and structure-breaking mutations and can be applied to arbitrary RPKI implementations without adaption. It also introduces new techniques for batching inputs, not considered in \cite{ammann2024dy}.

{\updated Recently, \cite{bennett2024ransacked} introduced ASNFuzzGen that translates ASN.1 definitions to code modules, allowing them to build on top AFL++ structure-aware mutation functionality to generate fuzzing inputs. This allows structure-aware fuzzing, but requires structure definitions and relies on existing mutation engines producing high-quality ASN.1 object mutations. Their approach cannot handle broken ASN.1 structure. Our work instead implements a template-agnostic ASN.1 parser not requiring ASN.1 definitions, can generate ill-formatted objects, and introduces a custom-built ASN.1 mutator which we show to outperform the general-purpose mutators.}

CURE \cite{mirdita2023cure} presented the first approach of black-box fuzzing RPKI software and uncovered a range of vulnerabilities, with the large majority of bugs (89\%) identified in parsing code. As we will illustrate in Section \ref{sec:challenges}, their approach has a range of limitations, requiring parseable objects, lacking coverage feedback, and no corpus. Our work overcomes these limitations by implementing an RPKI fuzzer that heavily utilizes coverage-guidance, a corpus of real-world objects and a range of other improvements that allow us to detect 21 new vulnerabilities overlooked by \cite{mirdita2023cure}. {\updated RFuzz \cite{shang2025fuzzing} recently presented new fuzzing work of RPKI using structured semantic mutations with context-free grammar to identify flaws in RPKI validators. The limited set of available mutations and lacking coverage feedback only allows them to identify 7 validation inconsistencies, overlooking all vulnerabilities discovered by this work.}

\section{RPKI Fuzzing is Hard}\label{sec:challenges}
RPKI fuzzing remains an open question in research. 

\subsection{Limitations of General Purpose Fuzzing}
\begin{figure}[t!]
    \centering
    \includegraphics[width=1.0\columnwidth]{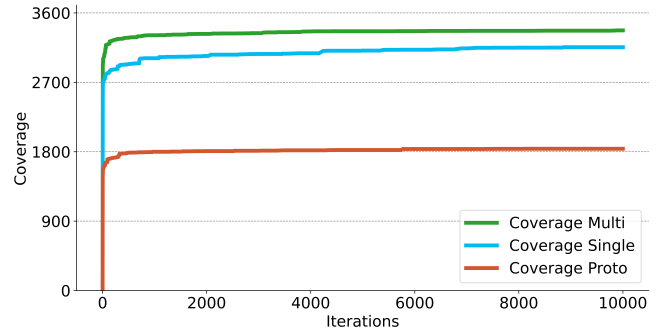}
    \caption{{\updated Branch} Coverage of ROA fuzzing with libFuzzer using built-in mutators with 1 and 10 objects, and with libprotobuf.}
    \label{fig:libfuzz_cov}
\end{figure}

General-purpose fuzzers were only applied on a very limited subset of RPKI validator functionalities. None of the existing fuzzing harnesses\footnote{\href{https://github.com/NLnetLabs/routinator/tree/main/fuzz}{github.com/NLnetLabs/routinator/tree/main/fuzz}} target deep or complex RPKI functionality, focusing narrowly on parsing routines and omitting validation logic entirely. To illustrate the challenges in applying existing general-purpose tools to RPKI, we create a structure-aware coverage-guided fuzzing setup using libFuzzer and the most popular RPKI validator Routinator. We build a custom harness for Routinator's validation routine and implement the required code adaptations to allow libFuzzer structure-aware fuzzing of the Routinator code. 
{\updated
For this, we use the libFuzzer wrapper for Rust called cargo-fuzz. 
We adapt Routinator and its libraries rpki-rs and bcder to include derived \textit{arbitrary} features on all necessary structs to allow libFuzzer to construct RPKI objects from raw bytes. For bytes which can't automatically derive the arbitrary trait, we implement a short routine that constructs byte objects from an array of u8 values. 

The fuzzing harness takes an RPKI object, like a ROA, as input, and nests it into a valid RPKI repository. For this, we create a custom tool that implements the respective required publication pipeline for any RPKI object type. 

\textbf{libprotobuf setup.} To additionally test the success of libprotobuf fuzzing, we also adapt the existing libprotobuf tooling \footnote{\url{https://github.com/google/libprotobuf-mutator}} to support mutations of structured RPKI objects. For this, we create protobuf templates for all RPKI objects that reflect the structure of real-world objects. We then adapt the existing DER encoder for protobuf objects\footnote{\url{https://github.com/google/libprotobuf-mutator-asn1}} to support the custom RPKI objects and encode them into valid DER. Finally, we hook a custom mutator into cargo-fuzz to use libprotobuf mutators instead of the built-in (byte-based) libFuzzer mutator for the objects. Within the harness, we use the DER encoder to encode protobuf bytes to valid DER.

For testing the two setups, we implement a fuzzing harness that executes a full validation run of Routinator, i.e., runs validation of an entire RPKI repository that contains the fuzzing input. Coverage results of this fuzzing are shown in Figure~\ref{fig:libfuzz_cov}, comparing libFuzzer with a 1 ROA repository, 10 ROAs (multi), and libprotobuf mutators.}
Clearly, libFuzzer structure-aware mutators outperform the libprotobuf mutators, which is in line with libprotobuf documentation recommending the use of built-in libFuzzer mutators. The evaluation also shows that, while the fuzzer initially finds new coverage, the speed of new coverage exploration quickly slows down. In total, we run libFuzzer for 3x8h but do not identify any new vulnerabilities, including none of the vulnerabilities of this work. We describe the reasons for this limited performance of existing general purpose tooling to fuzz RPKI next. 

\textbf{Cryptographic input complexity lowers fuzzing success.} 
RPKI objects contain a complex structure of inter-dependent fields and signatures for validation. For example, a ROA contains at least 51 separate field values, ranging from signature bits to key fingerprints and signing times. The validation of a ROA requires that all these fields have correct values, with at least 16 of the fields depending either on other fields within the same object (e.g., the key fingerprint needs to match the hash of the contained key), or on fields in other RPKI objects in the same repository (e.g., the parent fingerprint matching the hash of the parent key). If any of these fields are incorrect, the object fails validation and aborts the validation routine. This is problematic, as it will prevent the fuzzer from detecting any vulnerabilities in the code behind the first failed validation check. 

To ensure objects do not fail early in validation, the fuzzer should repair all fields that are not intentionally invalid before executing the target, e.g., re-computing signatures on the changed object content. However, inferring intentionally invalid fields is not possible with existing general-purpose tooling since they do not flag which bytes are intentionally manipulated and which could be overwritten. For example, the fuzzer might intentionally write manipulated bytes into the signature field, and correcting the signature in the harness would overwrite this mutation. Additionally, inferring where to insert corrected fields is not trivial, as the fuzzer might intentionally change object structure, e.g., switching locations of fields. These limitations make applying existing tooling to fuzzing cryptographic inputs challenging, which is especially true for inputs with complex cryptographic relationships like in RPKI that go beyond a single signature verification.

\textbf{Encoding limits input space.}
RPKI objects are encoded into binary using Distinguished Encoding Rules (DER). In our structure-aware fuzzing with libFuzzer, we rely on Routinator to encode the structured objects to valid DER. Accordingly, existing work generally uses general-purpose DER encoders for their fuzzing, which limits their fuzzing inputs to objects that can be validly encoded \cite{mansur2020detecting, ammann2024dy}. Alternatively, they use byte-level mutations, like havoc in AFL++ \cite{AFLplusplus-Woot20}, to manipulate the raw output bytes after encoding without considering structure. However, since this stage is not structure-aware, it relies on randomness over intentionally mutating, e.g., tags or field lengths, limiting its success with intricate complex object types, like in RPKI.

\begin{figure}[t!]
    \centering
    \includegraphics[width=1.0\columnwidth]{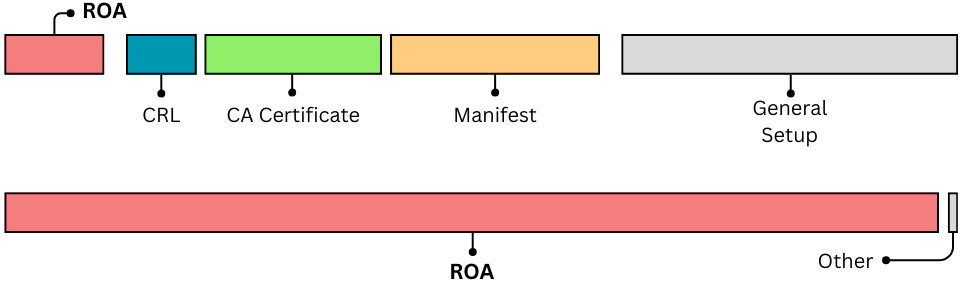}
    \caption{Flamegraph of Routinator validation with batch size 1 (top) and 1000 (bottom).}
    \label{fig:flamegraph}
\end{figure}

\textbf{Cryptographic setup slows fuzzing.}
An additional substantial limitation of RPKI fuzzing with existing tools is the complex cryptographic setup required by RPKI validation. For example, validating one ROA not only requires the object itself, but also a valid manifest, CA certificate, CRL, and {\updated Trust Anchor Locator (TAL)}. Without these objects, ROA validation can not be executed. The fuzzer thus needs to create a full repository setup for each fuzzing input, and the validator must validate this setup additionally to validating the actual fuzzing input. For our libFuzzer setup, we create a custom RPKI repository tool that can nest RPKI objects into a valid repository. However, repository creation and processing take computational effort, slowing down fuzzing significantly. To illustrate this, refer to Figure~\ref{fig:flamegraph}, which shows a flamegraph of Routinator when processing one ROA fuzzing input. Routinator only spends around 13\% of the total processing time on the ROA, with 87\% wasted on necessary validation overhead.  Multiple techniques exist to handle high per-input setup overhead, prominently snapshotting and forking \cite{schumilo2021nyx} which cache the setup state to speed up per-input fuzzing. In our setup, we snapshot a valid RPKI repository to nest inputs for faster fuzzing. Still, even with caching, libFuzzer can only test about 10 objects/s, illustrating these techniques only have limited success in RPKI. Snapshotting requires \textit{constant} setups for caching, but large parts of the RPKI setup are dynamic and depend on the fuzzing input. For example, the manifest contains a hash of the input, so this hash must be re-computed each iteration and the manifest must be re-signed in each run. Accordingly, multiple setup steps depend on the input and can not be cached. This leads to slow fuzzing speeds, even with snapshotting enabled. 

Overcoming this limitation can be achieved through batching many inputs into one setup and running this setup in one fuzzer execution. This distributes the setup creation/validation effort over multiple inputs and significantly lowers per-input overhead. The effect is shown in Figure~\ref{fig:flamegraph}. When 1000 ROAs are batched into one setup, per-input overhead is reduced by 66x and Routinator spends over 99\% of processing time on actually processing the ROAs. Batching has the additional advantage of testing more realistic input configurations: Real-world RPKI repositories, on average, include 6 ROAs. Testing multi-object batches captures this real-world complexity and can, e.g., also detect issues in processing of multiple similar ROAs with overlapping or conflicting resources. The benefit of batches for more coverage is evident in Figure~\ref{fig:libfuzz_cov}, with the multi-input approach outperforming the single input fuzzer by roughly 7\%. 

As we show in Section~\ref{sc:non_seq}, batching significantly improves fuzzing success and speed for RPKI, but comes with its own limitations regarding coverage-guidance.

\textbf{Challenges of batch fuzzing.}
Existing tools are designed for sequential fuzzing: they mutate and run one input, check performance, then move to the next input. This design is necessitated by how the tools use coverage. If a batch of multiple inputs is tested per run, the sequential tool only rates the performance of the entire batch. 

For example, consider libFuzzer executing the RP with a batch of 10000 objects, and discovering 100 new branches. It will rate this input as well performing. However, in a worst case, a single object is responsible for 100 new branches and 9999 objects performed poorly. Since libFuzzer can not determine which part of the input performed well, it rates the entire batch as well-performing, despite only 0.01\% of the input yielding new coverage. Our experiments show that in most runs, only a few objects of a batch perform well, making sequential coverage-guidance immensely inefficient. 
Previous work  on TLS fuzzing encountered this issue when testing TLS certificate chains \cite{chen2015guided}. To fuzz the complex interaction within a chain, \cite{chen2015guided} construct a batched certificate chain from multiple mutated objects. However, after running this chain, they can only extract coverage results of the {\em entire} batch, which does not allow pinpointing which of the certificates resulted in a coverage increase. To overcome this problem, \cite{chen2015guided} only adapt a single certificate in the chain for the next iteration and then again check the coverage. While this improves accuracy, it also significantly slows down the fuzzer, as previous inputs are tested again without adaption, thereby not yielding new results. Consider a certificate chain of five certificates. If only a single certificate is changed in each iteration, the fuzzer spends 80\% of its time testing already used certificates that do not provide new insights. When even larger batches are used, e.g., to test more complex certificate chain structures, the slowdown gets increasingly worse. This illustrates a fundamental trade-off in fuzzing complex cryptographic protocols with current tooling: larger batches of inputs allow testing complex interactions (like multi-ROA repositories) and substantially speed up per-input fuzzing (by up to 66x), but introduce immense slow-downs to overall fuzzing speed due to the inherently sequential design of current coverage-guided tools. Since fuzzing success relies on coverage-guidance \emph{and} testing as many objects as possible, this constitutes a major hurdle and has led many works on fuzzing cryptographic protocols to not use coverage-guidance at all \cite{mirdita2023cure, somorovsky2016systematic, brubaker2014using}.

CAT overcomes these issues through novel techniques to score individual objects within a batch, explained in Section~\ref{sc:non_seq}.

\subsection{Limitations of existing RPKI Fuzzing}
Due to the limitations of existing tooling, one previous work has developed a purpose-built fuzzer for RPKI called CURE \cite{mirdita2023cure}. However, CURE includes many issues of existing general purpose fuzzers and does not find deep and complex vulnerabilities in RPKI validators due to several technical and conceptual limitations: 

\indent {\bf Lack of coverage feedback.} The fuzzer in \cite{mirdita2023cure} is black-box and does not utilize any coverage feedback from the tested software. Their method relies purely on random inputs to discover vulnerabilities. In a complex validation setup like RPKI, this limits the fuzzer in finding deeper nested vulnerabilities and more complex bugs due to the lack of insight into the program's internal state \cite{takanen2018}. As a result, the fuzzer gets stuck at the first parsing and validation functions.
Comparing the methodology of \cite{mirdita2023cure} and their open-source implementation against our setup, we find that it, on average, discovers 47\% fewer new code paths than \cure\ due to its black-box nature and limitations in object signing. CURE does not identify any vulnerabilities found in this work.

{\bf No corpus approach.} The methodology in \cite{mirdita2023cure} 
generates each object from scratch, prohibiting using a corpus. Previous research \cite{bohme2016} has demonstrated that a corpus of real-world objects can significantly improve branch coverage, thereby helping to find more vulnerabilities. Our approach incorporates a large corpus of real-world RPKI objects, improving the ability to discover vulnerabilities by starting with diverse inputs.\\
\indent {\bf Separation of object generation and signing.} We find several drawbacks in the design of \cite{mirdita2023cure}, which separates object generation and the RPKI functionality, including object signing. This separation requires parseable object structure to allow for post-mutation parsing and signing. If the objects are not parsable in \cite{mirdita2023cure}, signatures are not corrected and objects fail too early in the validation pipeline, limiting the fuzzer's effectiveness. This problem in \cite{mirdita2023cure} is evident in the lack of vulnerabilities found in the RP implementations using OpenSSL for parsing (Fort and rpki-client), since they are more resilient to parsing bugs. 
Consider, for example, a bug discovered in our work, where the tag of a field has an additional preceding "more bytes follow" byte, leading a validation check to throw a critical error. This bug could not have been discovered with a ill-structured object in \cite{mirdita2023cure} since the size of the object is increased by one, making it unparsable and therefore unfixable for \cite{mirdita2023cure}, leading to the object failing during the parsing stage of the RP before the bug could be triggered. A well-structured object also does not trigger this bug since it keeps the tags correct to secure parsability for signing. 
By integrating mutation and signing into a shared module, we achieve that fields are continuously repaired to maintain object validity. We further use the new technique of labeling fields after parsing {\updated so that} mutation does not prevent signing. This integration allows our fuzzer to handle non-parsable objects effectively, allowing them to proceed further in the validation pipeline and uncover more complex vulnerabilities \cite{bohme2016}, including vulnerabilities in Fort and rpki-client.

\indent {\bf ASN.1 object mutation.} \cite{mirdita2023cure} uses a generation-based approach with a manual schema definition. This simplifies tooling, as an existing ASN.1 mutation generator is used to create test objects instead of building a new or heavily adapting an existing ASN.1 mutation tool to support RPKI objects. However, this method results in a lack of diversity, as the singular structure definition is the only valid template the fuzzer has available. 
We present a novel approach for ASN.1 object generation that allows for structure-aware mutations without manual structure definition, increasing object diversity and applicability of our module to other ASN.1-based protocols. This also allows us to define ASN.1 type specific mutations to further increase mutation diversity.

\section{Non-sequential Coverage-Guided Fuzzing}\label{sc:non_seq}
Sequential single-input fuzzing is not successful in RPKI, lacking support for multi-object repositories and failing to handle the high per-input overhead effectively. This strongly limits applicability of all available tooling, and has led to large parts of the RPKI validation logic remaining untested. 
Next, we illustrate our novel conceptual contributions to overcome these hurdles, allowing us effective coverage-guided non-sequential fuzzing of RPKI.

\subsection{Extracting the coverage}
An essential prerequisite for coverage guidance is collecting runtime coverage of validators. For this, we utilize AFL++ coverage instrumentation, which is available for the three most popular RPs Routinator, Fort, and rpki-client. We compile all RPs with afl-clang compiler, which inserts execution counters into the compiled binary, tracking which branches of the program are executed during runtime inside a \textit{counter bitmap}. We then create a new standalone tool called \textit{cat\_coverage} that attaches to the binary and reads the coverage bitmap from shared memory. The tool continuously samples the bitmap and tracks overflows of counter values, which is necessary since counters are only tracked in a single byte and values over 255 could otherwise not be accurately measured.
We implement several optimizations to achieve sufficient speed to track all wrapping counters. Our optimizations include mapping the maximum counter index and only reading until this index, skipping counters which have already been scored in previous fuzzer runs, multi-threading, and using the \textit{unsafe} keyword in our rust code to implement faster shared memory access. With these optimizations, our tool samples the bitmap once every 2$\mu$s (Fort) - 150 $\mu$s (Routinator), depending on the bitmap size.

\begin{figure}[t!]
    \centering
    \includegraphics[width=1.0\columnwidth]{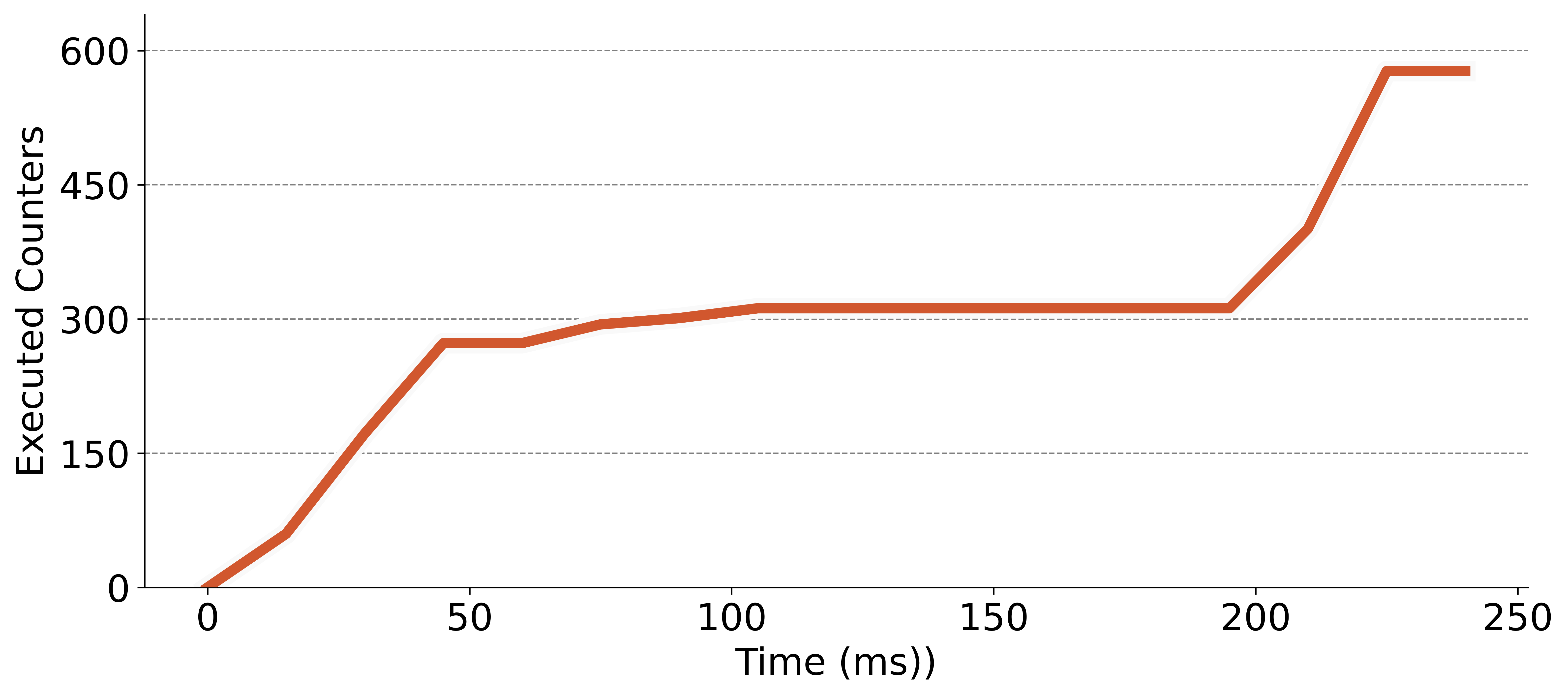}
    \caption{Function coverage progression.}
    \label{fig:cov_prog_base}
\end{figure}

\subsection{Tracking coverage with Coverage Progression} 
Tools like AFL++ read the final coverage counters after execution to check whether new branches were discovered during the execution. When running batches of objects, e.g., 1000 manipulated ROAs, the final bitmap only contains the summary of branches explored by the entire batch. The fuzzer cannot distinguish whether only one or a few objects within the batch resulted in new coverage.\\
\indent To overcome this problem, we propose the novel technique of Coverage Progression (CP). Instead of only using the final result (bitmap), like AFL++, CP continuously samples snapshots of the counter bitmap during execution, tracking \textit{when} counters are incremented. Figure \ref{fig:cov_prog_base} illustrates the CP of Fort while validating 1000 well-formatted ROAs. As expected, the graph shows that the amount of executed unique counters continuously increases during execution. The amount of new counters in the middle part of the runtime, i.e., between 48ms and 200ms, remains relatively stable. Manual investigation shows that during this period, the RP validates the collected ROAs, and many validations do not explore new coverage. Thus, the CP already provides an indication of which ROAs in the batch resulted in new branches; if a new branch was discovered in a bitmap snapshot at 55ms, it was likely caused by one of the first ROAs, while a branch discovered at 190ms would likely map to one of the last ROAs in the input. This time-based approach allows a rough mapping between coverage and objects and is thus already an improvement over the technique used by AFL++, but it is insufficient for a precise scoring. Some fractions of the stable section between 48ms and 200ms are not spent on ROA processing. Thus, a branch discovered at exactly half the interval (124ms) does not map to the middle ROA. We find the error for this approach is up to 30\% of the batch size, depending on the RP, which corresponds to a precision of 300 objects for a batch of 1000 ROAs.

\subsection{Tracking Objects with Identification Functions}
\indent To improve the precision of our technique, we propose the concept of Identification Functions (IFs). IFs serve as a side-channel into the execution of the target to leak which object from the input batch is processed when a snapshot is taken. Consider the following pseudo-code, which represents the loop to process a batch of (manipulated) ROAs. 
\begin{lstlisting}[style=clean]
for roa in input_roas do
    trav_roa(roa)             

function trav_roa(roa: Roa):
    % <--- use a counter here
    parse_roa()
    validate_signature()
    validate_fields()
    ...
\end{lstlisting}
The \textit{trav\_roa()} function is executed once per ROA. By using the coverage counter at the beginning of this function as a side-channel, we can identify which ROA is processed at any point. For example, if a batch of 100 ROAs is processed and the counter in the beginning of \textit{trav\_roa()} has a value of 42 in a CP snapshot, the 42th ROA was processed when the snapshot was taken. The IF side-channel thus allows precise coverage attribution in batches: If CP finds new coverage in a bitmap snapshot, the fuzzer can read the IF value to map which input object resulted in this new coverage. We evaluate accuracy of this technique in Section~\ref{subsec:accuracy}.

\subsection{Identifying suitable IFs}
A straightforward option to obtain IFs is manually inserting hooks into the processing of the target. However, a manual approach incurs substantial manual effort in identifying processing flows and loops and thus requires expert knowledge of the tested software. Instead, we propose an algorithm to automatically identify suitable IFs in arbitrary targets, significantly lowering the manual effort in fuzzer setup.
To find suitable IFs, we execute the RP with multiple batches of different sizes and check which functions have execution counts with a linear correlation to the amount of objects in the batch. The steps of IF identification are shown in the following pseudo-code. 
\begin{lstlisting}[style=clean]
filtered := get_all_rp_counters()
test_sizes := [20, 100, 200, 400]
for n in test_sizes do
    create_repository(n)             
    run_rp()
    counters := get_coverage()
    candidates := filter(x -> x == n, counters) 
    filtered := intersection(filtered, candidates)
identification_functions := filtered
\end{lstlisting}

In each IF identification run, we setup a repository with a fixed amount of manipulated objects, run the RP, then read the coverage counters and filter for counters that match the number of objects in the test-batch (line 7). IF identification must use \textit{manipulated} objects {\updated so} IFs also apply when objects are not well-formatted. The target is executed multiple times with different batch sizes to remove false positives, always removing all counters that do not match the number of objects. 4 iterations with different batch sizes always yielded no false positives in any RP.
\begin{figure}[t!]
    \centering
    \includegraphics[width=1.0\columnwidth]{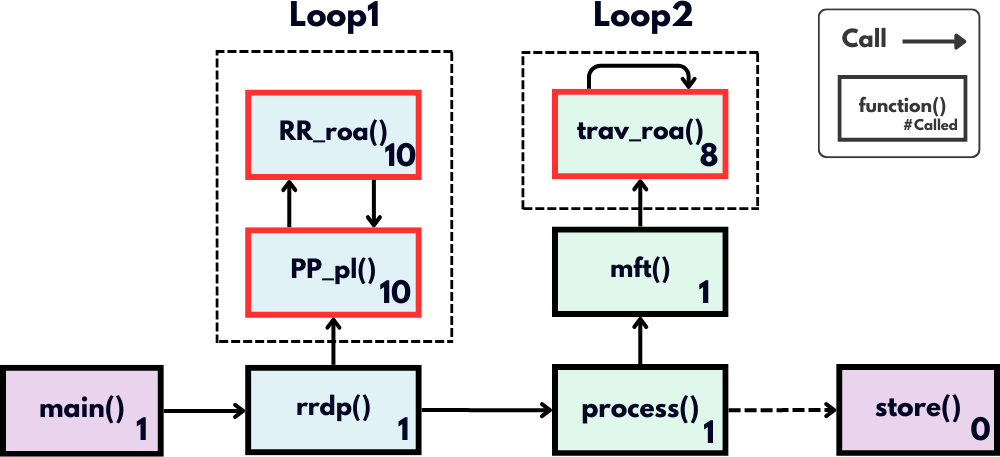}
    \caption{Example control flow of RP.}
    \label{fig:marker_three}
\end{figure}

\begin{figure}[t!]
    \centering
    \includegraphics[width=1.0\columnwidth]{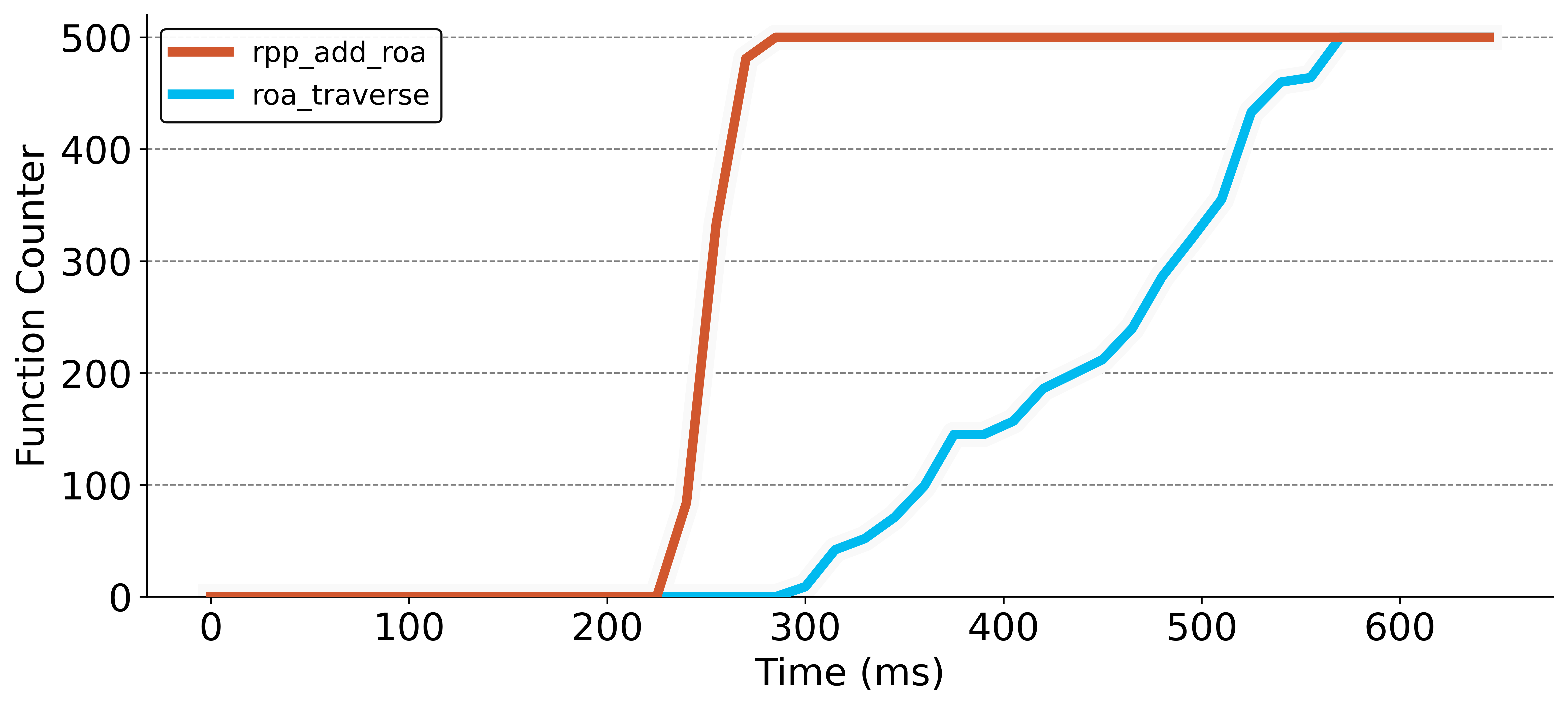}
    \caption{Execution counts of identification functions.}
    \label{fig:marker_cov}
\end{figure}

\textbf{Dealing with multiple loops.}
The processing of objects does not always occur in a single loop. Consider Figure \ref{fig:marker_three}, which shows an abstraction of the function graph of Fort. In total, three functions are identified as IFs (red border) using the above technique. These functions are called at different stages of the execution. The first two functions are executed during download, while the function \textit{trav\_roa()} is executed during the validation phase. If only one of these functions is used as IF, e.g., \textit{pp\_pl()}, a new branch discovered during ROA validation could not be attributed.\\
\indent To overcome this, we use all identified IFs, utilizing each function as the indicator for their respective part of the execution. For example, if a new branch is discovered during download, \textit{RR\_roa()} will give the best indication of the current object, while a new branch during validation would better be mapped with \textit{trav\_roa()}. We use the following decision algorithm to decide which counter is best suited as IF at a given sample snapshot, where $C_{I}$ describes the list of counter values for all IFs, $B_s$ refers to the batch size, and $e$ to individual counters. The counter value is given as: $v = max([e \in C_{I} | (e < B_s \wedge e \ne 0)]$. The idea behind the formula is that the currently executed IF in the loop will generally have a value which is neither 0 nor batch-size, while the other loops will either already have finished (i.e., counter == batch-size) or not started yet (i.e., counter == 0). We use the max value here to {\updated find the latest function already executed within a loop,} as the started execution of a given object could already have triggered the new branch.\\
\indent We illustrate the use of IFs with a batch of 500 ROAs in real-world Fort in Figure \ref{fig:marker_cov}. 
As visible from the graph, the two functions \textit{rpp\_add\_roa()} and \textit{roa\_traverse()} describe separate loops; the second function is executed for the first time after the first function finished execution.
Further, parsing and validation takes longer than the download processing of ROAs. Using these two IFs, we can map the discovery of a new branch to a specific object, illustrated in the following pseudo-code. Each time a new branch is found, IF values are extracted and used to map the increase to a specific object. 

\begin{lstlisting}[style=clean]
while rp_thread.alive() do
    counters := get_current_coverage()
    new_counters := get_new_counters(counters)

    if new_counters.size() > 0 then
        if_values := counters[ifs]
        b_position := selection_formular(if_values)
        known_counters.extend(new_counters)
\end{lstlisting}

\subsection{Evaluation of IF Accuracy}\label{subsec:accuracy}
\begin{figure}[t!]
    \centering
    \includegraphics[width=1.0\columnwidth]{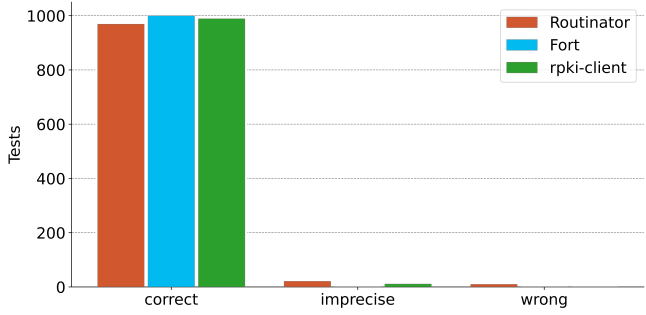}
    \caption{Comparison accuracy of IFs in RPs.}
    \label{fig:if_comparison}
\end{figure}

To test the accuracy of IFs in mapping coverage increases to objects in large batches, we setup Routinator, Fort, and rpki-client in a local testbed. We create repositories of 1000 ROAs, and repeatedly insert a manipulated ROA at a random position within the batch. We then execute the RPs while running CP with IFs and check whether the position of the manipulated ROA in the batch is correctly inferred. We score the results into \textit{correct inference}, \textit{imprecise inference} (the coverage was attributed to the correct object but also the previous or next object), and \textit{incorrect inference} (the wrong object was attributed to have caused the coverage). Results are presented in Figure~\ref{fig:if_comparison}, showing high accuracy in all RPs. The accuracy depends on the size of the coverage bitmap: the larger the bitmap size, the longer the tool takes to snapshot the map each iteration, lowering the sampling frequency. For Routinator, the large bitmap (402052 bytes) takes around 150$\mu$s to sample, while Fort (8257 bytes) only takes around 2$\mu$s.\footnote{This size difference stems from Routinator implementing parsing/validation from scratch, and offering additional services like a UI.}
Inference errors occur if new coverage is triggered at the end of function execution, between bitmap snapshots. Then, the new coverage will only be detected in the next bitmap snapshot which already includes the next execution of the IF, so the coverage will be wrongfully attributed to the next object, {\updated yielding an  \textit{imprecise} result. If more than one object was processed between samples, the coverage attribution becomes wrong.} The higher the sampling frequency, the smaller this window of error. With our optimizations, the sampling frequency is sufficient for a low margin of error, showing 1\% error in Routinator, and no errors in Fort and rpki-client. 

Importantly, this high accuracy requires sequential processing of objects in the target software, as out-of-sequence processing breaks the direct correlation between IF counter and index in the batch. Both rpki-client and Fort include a feature to randomly shuffle ROA processing order, which we disable during fuzzing to allow IF mapping. None of the RPs use multi-threading during object processing, which would also interfere with IF accuracy. {\updated In total, using CP with IFs improves precision of identifying the correct object in a batch of 1000 objects from  1/1000 correct objects (0.1\%) for regular libFuzzer coverage and 30/1000 correct objects in the timing approach (3\%) to 990/1000 correct objects (99\%) in CAT.} CP with IFs thus enables effective coverage-guided fuzzing with object-specific coverage attribution even in large input batches. 

\section{CAT Architecture}\label{sc:architecture}
\begin{figure}[t!]
    \centering
    \includegraphics[width=\columnwidth]{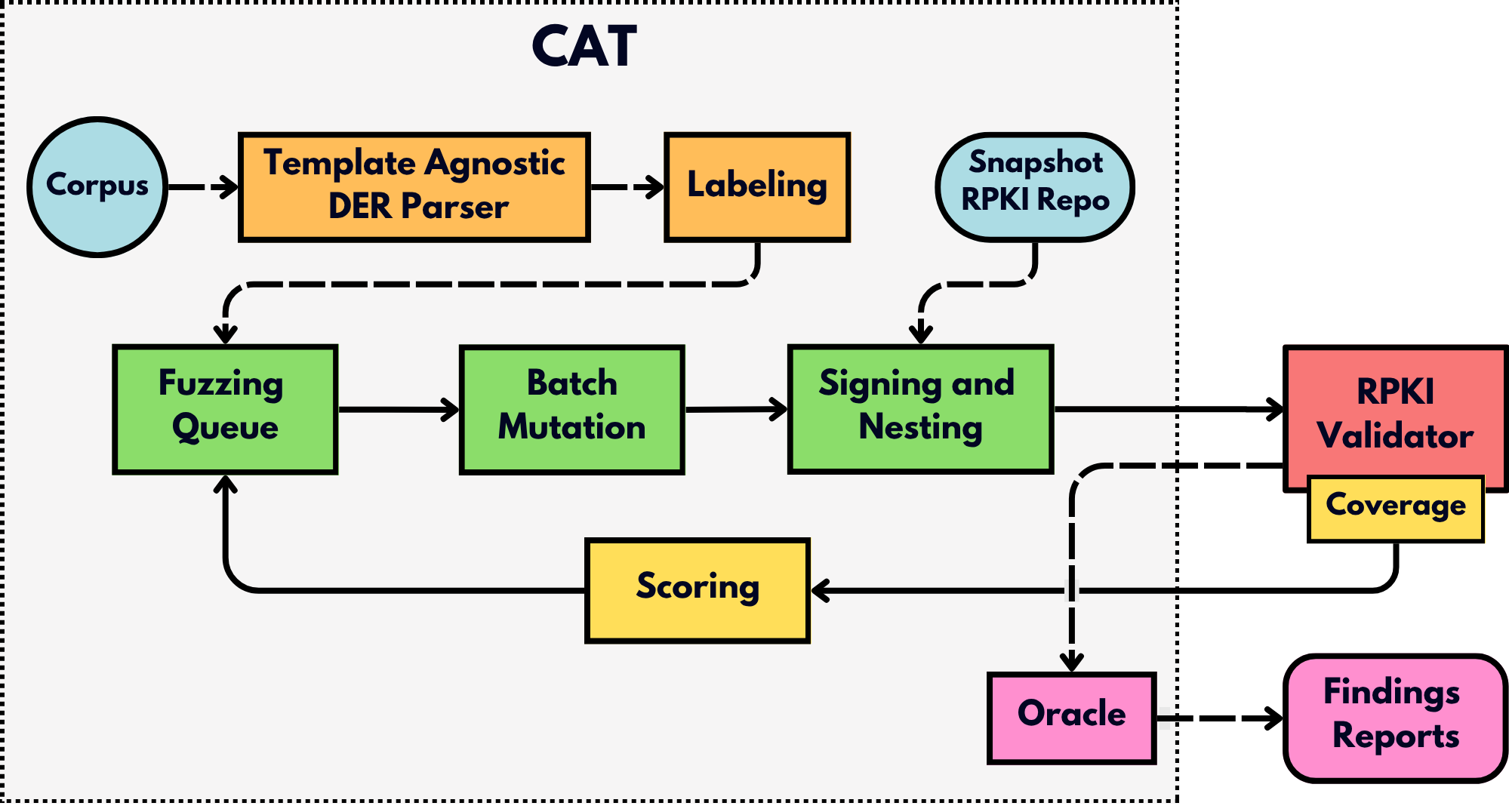}
    \caption{Architecture of \cure\ with multiple modules.}
    \vspace{-10pt}
    \label{fig:curepp_architecture}
\end{figure}

To illustrate the benefits of our developed coverage-guided techniques, we incorporate them into a powerful fuzzing tool called CAT. The architecture of CAT additionally includes new contributions in properly handling the complexity of RPKI's cryptographically linked input model and encoding requirements, which culminates in the following core innovations: 
a template agnostic DER parsing and encoding module, a ASN.1 mutation engine capable of preserving validity and re-signing after complex structural changes, and a batch aware feedback loop that utilizes coverage progression and identification functions to maintain guidance at scale. Combined, these components enable high-throughput, structure-aware fuzzing that systematically exercises both parsing and validation stages, making \cure\ a powerful and generalizable approach for testing not only RPKI, but other cryptography-based Internet protocols. 
Figure \ref{fig:curepp_architecture} shows the architecture of \cure. 

\subsection{Queue and Corpus}
\cure\ maintains a fuzzing queue of candidate RPKI objects for mutation. At initialization, the queue is seeded with a balanced mix: 50\% are drawn from a corpus of real-world objects collected from a Routinator validation run and randomly selecting a subset of objects that is half of the targeted batch size, and 50\% are from a self-generated corpus of well-formatted objects. This split proved effective in practice, combining the structural diversity of real-world data with a high likelihood of producing valid test cases after mutation. Prior to insertion, each object is parsed into an ASN.1 syntax tree using our custom DER parser.
The parser requires no ASN.1 schema and can handle any DER-encoded RPKI object type, mapping each field and its relationships in the tree structure. This enables structure-aware mutations and the automatic repair of dependent fields. The fuzzer additionally implements a labeling engine that uses object structure to name each field in the tree. Labels are constructed from the respective RFC that define the structure of each RPKI object. The labeling logic is an important contribution as it allows the signing logic to locate and update all fields affected by a mutation, even when the object’s structure changes substantially, which is not possible with existing tooling. Labeling is optional, enabling fuzzing of arbitrary ASN.1 objects without predefined structure. However, without labeling, the fuzzer is not able to repair cryptographic fields after mutation.

\subsection{Batch Mutation}
The mutation module manipulates objects to trigger crashes and unexpected behavior in the RPs. Based on the shortcomings we identified in \S\ref{sec:challenges}, we develop a new structure-aware mutation logic unifying the concept of random byte mutations, which we find more suited to discover parsing bugs, with structure-aware mutations, which are targeted to find bugs in the validation code.\\
\indent \textbf{ASN.1 tree mutations.}  
\cure\ implements a new mutation logic, based on the ASN.1 syntax tree of the RPKI objects. Existing applications of ASN.1 syntax trees for mutations, e.g., \cite{wang2023asn}, are limited: they are slow, require an ASN.1 schema for object generation, and mutations are limited to only a few operations that keep a correct syntax {\updated so that the} object can still be encoded by an existing ASN.1 library. In contrast, we propose a new syntax-tree based mutation logic that implements 59 different mutation options for ASN.1 fields, ranging from string manipulations to date operations. Using our purpose-built ASN.1 encoder / decoder ensures that all mutated objects can still be processed, making our tool independent of the capabilities of existing ASN.1 encoders. When executing the mutation module, \cure\ selects a random node within the tree for manipulation. The mutator then chooses one of the following:\\
\indent \textit{Structure-level mutation:} Mutate the ASN.1 structure by manipulating the length field or the tag value.\\
\indent \textit{Type-level mutation:} Mutate the data of the node according to its type. For example, a string value will receive string-based mutation, like capitalization of random characters or insertion of interesting characters, e.g., null-bytes.\\
\indent \textit{Byte-level mutation:} Mutate the node data randomly without considering any structure-aware data. Byte-level mutations include insertion of random bytes, bit-flips, or removal of byte-sequences.\\
Note that type-level and byte-level mutations lead to objects with a valid ASN.1 structure and are thus generally parsable by the RPs. This allows us to test deeper parsing code, like the parsing of the content of fields, and the validation code behind object parsing. Further, using the tree-based approach allows us to sign and encode manipulated objects with our custom DER-encoder, even if they are not parsable.\\
\indent To illustrate the benefit of our approach, consider a vulnerability we found that is triggered by additional bytes in the subject public key field. Adding bytes to this field makes the object not parsable, since the length value of this field is wrong and thus, \nv{fuzzing harnesses} that require parsability to further process the object, like for signing, cannot handle this object and would therefore feed it as-is to the RP. The RP will fail parsing at an early stage since the length is wrong, and not even reach the stage of parsing the content of the subject public key field. The vulnerability is not detectable.\\
\indent In contrast, using the tree-based approach, the mutator marks the field's manipulated content and field lengths. This allows the encoder within the harness to repair the field length, thereby making the object parsable for the RP. The RP reaches the code that parses the field content and crashes as the vulnerability is triggered.\\
\indent Further, we found implementing type-level mutations provide a substantial benefit over random mutations. As edge cases for specific types are more likely to trigger bugs, implementing type-level mutations speeds up bug detection. For example, the mutator used 0x1F as a known interesting tag, representing a value of \textit{more-data-follows}, which triggered a vulnerability in Fort that did not consider the existence of this tag. Likewise, another bug in Fort was triggered by sequence mutation, deleting a specific element from the encapsulated content sequence. 

\textbf{Splitting and splicing.}
In addition to the mutation of nodes within the tree, we implement splitting and splicing, used by fuzzers like AFL++ or libFuzzer \cite{AFLplusplus-Woot20} to combine parts of different objects into a single object to create new diverse inputs from an existing corpus. We implement this concept by combining different ASN.1 trees into a single tree. For this, a random node from the tree of the object is selected, deleted and replaced with a node from the other tree, including all of its children. 
We implement this algorithm to also allow combination of different object types. For example, the mutator might insert the NameAndHash list from a manifest into a ROA, to test how RPs react to unexpected field types in a given ROA. This approach allows us to create more diverse and complex objects, as well as to create objects with unexpected but valid ASN.1 structures. 

\subsection{Signing, Encoding and Nesting}
After mutation the manipulated objects are fed to the RPs and \cure\ performs post-processing {\updated so that} each object can pass initial parsing and reach the code handling the mutated fields. This is essential for RPKI fuzzing, as even minor structural breakage would otherwise cause early rejection and prevent deep validation logic from executing.\\ 
\indent \textbf{Structure repair.} We use a taint-based mechanism to allow the harness to identify which parts of an object require DER structure repairs. If an object field receives a structure-impacting mutation, like mutations that implicitly change the length of a field, the field is marked as tainted. The taint further propagates iteratively to all parent nodes of this field, as a length change of a child also always results in a length change of the parent. For example, consider a sequence of 20 bytes, containing two children of 10 bytes each. If the content of the first child is manipulated and the length changes to 12 bytes, the child length field has to be adapted to 12 bytes, but the length of the parent also has to be adapted to 22 bytes. The harness thus checks the length values of all tainted fields, starting with the deepest children and iteratively adapts the length of the tainted parents.\\
\indent \textbf{Field repairs.} RPKI validation requires some (cryptographic) object fields to be correct to further process the object. The harness needs to adapt these fields so processing does not fail before the mutated object part is processed. To achieve this, the field labels are used. The harness implements all necessary changes to fields for each RPKI object type to fix object validity. Intentionally manipulated fields are {\updated flagged to prevent overwriting mutations.} This flag can also be applied to structure elements, like lengths, to ensure the fixing does not overwrite structure-impacting mutations. Using labels allows the harness to fix nodes even if the object is not parsable, or fields have been moved to another location by the mutator. This is a main advantage over, e.g., relying on a template to find the necessary fields, as the template would fail in most mutated objects. After fixing all the necessary fields, like object signatures and hashes, each object is encoded in our custom encoder with the DER-encoding rules, as dictated by the RPKI standard. The encoder tracks all structure-breaking mutations of the object, like changes in lengths or ASN.1 tags. To still encode such broken objects, the encoder first encodes the object into DER without these mutations, then applies these breaking mutations to the output bytestream after encoding. This allows our custom encoder to encode objects with broken ASN.1 structure, which is not possible with other current tooling. \\
\indent \textbf{Nesting.} After fixing and encoding the objects, the batch of objects is nested into a full RPKI repository that includes all required auxiliary objects. This repository is loaded from a cached snapshot to improve efficiency. After inserting the objects, all required cryptographic fields are updated in the auxiliary objects, like fixing manifest hash and signature, and re-computing the RRDP snapshot hash. After creating the repository structure, \cure\ creates the required files to allow the RPs to fetch the repository over RRDP and rsync.

\subsection{Target RPKI Validators}
After creating the repository, \cure\ executes the targets. In our setup, the targets include all RPs used in real-world RPKI deployments, including Routinator (0.14.0), Fort (1.6.2), rpki-client (9.1), OctoRPKI (1.5.10), and Prover (0.9.2). Note that OctoRPKI was deprecated in June 2024; we include it in our study since it is still used by many real-world systems.\\ 
\indent We implement coverage-guided fuzzing in all RPs that provide the required compiler support for the coverage-instrumentation, which includes Routinator, Fort, and rpki-client. All three RPs are compiled with coverage-instrumentation enabled. We implement \cure\ to interact with the RPs over a configurable interface, defined by one JSON file per RP. Adding an additional RP to the setup only requires creating a new JSON file that lists the call parameters of the RP, like the flag for the repository folder or the binary location. All RPs are executed in parallel to speed up fuzzing. \cure\ waits for all RPs to finish processing prior to moving to the scoring and oracle modules.

\subsection{Coverage and Scoring}
When executing the validator, CAT additionally attaches the coverage module to the target through a shared memory segment. This allows the coverage module to continuously track CP/IFs and map coverage increases to specific inputs. After execution, per-object coverage is read from the coverage module and used to score each mutated input individually:  
We use the amount of new branches as the score for the object, limiting the maximum score to 10, which showed a good trade-off in practice between prioritizing well-performing objects while not over-emphasizing single objects in the batch. Objects that did not discover new coverage are discarded for consecutive runs. 

\subsection{Oracle}
The oracle module is only run after RPs finished processing, and determines whether the inputs triggered unexpected or anomalous behavior. If such behavior is detected, the oracle module triggers result logging for further triage. The oracle uses 4 metrics to detect unexpected behavior:\\ 
\indent \textbf{Crash detection.} If an RP crashes, \cure\ logs a new finding in a short report, detailing the batch that triggered the problem, and the output of the crashing RP. We detect crashes via a lack of output from the RP, or crashing key-words in the logs.
This approach does not introduce any false-positives or -negatives. For Prover, which runs as a server process, \cure\ also checks if the process has terminated unexpectedly.\\
\indent \textbf{Execution stalls.} \cure\ additionally measures the processing time of each batch and flags runs exceeding twice the baseline execution time (average of five valid batches before fuzzing). This detects performance degradation or infinite loops without introducing false positives.
If the threshold is surpassed, the unexpected delay is logged and a report is created, including the RP log and the execution time. This oracle detects long execution and stalls.\\
\indent \textbf{Validation integrity check.} In each iteration of the fuzzer, we include an additional independent CA with a test ROA. This ROA should always be included in the validated output of the RP, as it runs separately from all testing CAs. If the test ROA is not included, a breach of the separation of the validation between different CAs is logged in a report, including the log output of the respective RP.\\
\indent \textbf{Cross-implementation consistency.} After each iteration, the fuzzer additionally compares the accepted ROAs of each of the RPs. If validation results do not match, this indicates that the RPs reached inconsistent results on the validity of at least one object. The observation is logged to identify validation differences between different implementations.

Every time the oracle finds an issue in the target, like a crash, a report is generated that contains information about the manipulated inputs, the oracle that identified the issue and the log outputs of all RPs.

\section{\hspace{1mm}Evaluations}\label{sc:eval}
We evaluate the performance of \cure\ along three dimensions: 
(1) We test raw throughput to illustrate the overall fuzzing speed \cure\ provides.
(2) We evaluate the effectiveness of our new techniques in achieving high coverage by comparing coverage results against setups with baseline fuzzers.
(3) We measure the individual contribution of each technical design component of \cure\ via ablation.

Experiments are run on a current generation Intel i7 (with 8 cores, 16 threads, 32 GB RAM) running Ubuntu 22.04, fuzzing five RPKI relying parties (Routinator, Fort, rpki-client, OctoRPKI, Prover) under identical conditions and all fuzzers use the same initial seed corpus. 
\subsection{Fuzzing speed} 
With our optimizations, \cure\ generates around 2000 objects/s, which surpasses the processing speed of the fastest RP Routinator (1500 objects/s). The maximum fuzzing speed is thus not limited by \cure\ but by how fast the RPs process generated objects. The slowest RP Prover processes around 730 objects/s, limiting the overall fuzzing speed of our setup when testing all RPs simultaneously to 730 objects/s.\\
We run the fuzzer on each object type for 2h. We chose 2h since no new branches were discovered within 1 million test objects after 2h for any type. For reliability, we re-start the fuzzer four times for each type to ensure no missed branches due to the fuzzer getting stuck in some code-branch, resulting in 8h test time per type. In total we run the fuzzer for 48h to test all object types, resulting in a total of 300 million test objects within a timespan of 7 days.
\begin{figure}[t!]
    \centering
    \includegraphics[width=1.0\columnwidth]{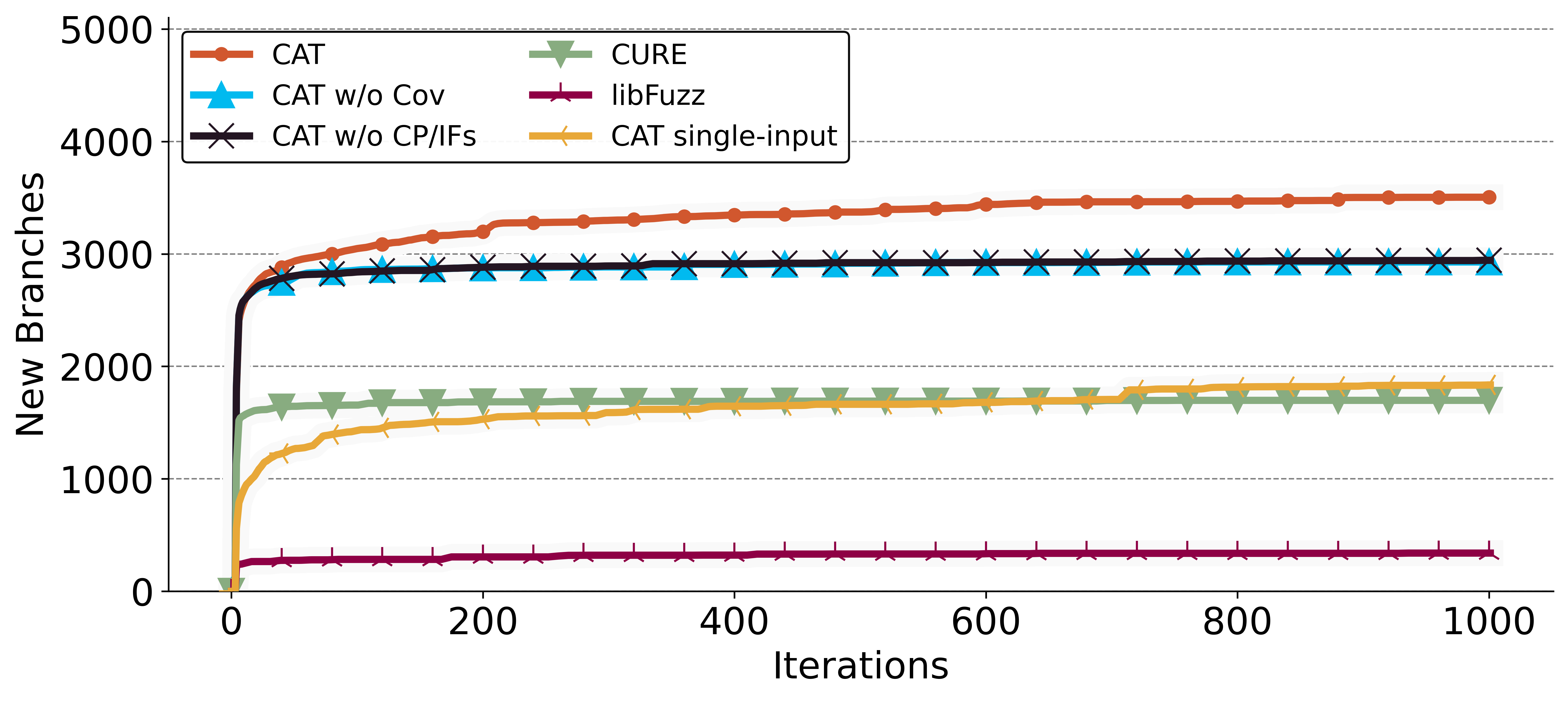} 
    \caption{Branch coverage increase.}
    \label{fig:cov_inc_all}
\end{figure}

\subsection{Comparison to previous Work} 
We evaluate the benefit of our new techniques in CAT using different configurations:\\
\textbullet\ \cure: Full coverage-guidance using coverage-progression and identification functions.\\
\textbullet\ Batch Coverage: CAT without object-specific coverage scores, not using CP/IFs.\\
\textbullet\ No Coverage: CAT without coverage-guidance\\
{\updated \textbullet\ No Batch: CAT without batching}\\
\textbullet\ CURE: The only other open-source available RPKI fuzzing tool CURE \cite{mirdita2023cure}, which lacks coverage-guidance and custom ASN.1 mutation and labeling modules.\\
\textbullet\ libFuzzer: State-of-the-art libFuzzer with structure-aware mutations, snap-shotting and multi-object support for batches.

The results are illustrated in Figure \ref{fig:cov_inc_all}, plotting the improvement of branch coverage over a baseline of a 1 ROA repository without manipulated objects. 
In the test setup, we ran batches of 1000 manipulated ROAs against RP implementations, observing similar performance across all the RPs.
{\updated To isolate the effects of batching, we include a configuration of \cure\ with a single-input where each execution processes only one object while retaining all other components (mutation, repair, scoring). This configuration enables direct comparison between batched and non-batched execution under identical conditions.
In particular, batching improves coverage growth and reduces time-to-first vulnerability, which is beyond pure throughput gains.}
The figure clearly illustrates that \cure\ outperforms \cite{mirdita2023cure} and libFuzzer in all configurations, showing the increased fuzzing success due to the improvements in the fuzzing architecture, most notably the object mutation logic through syntax trees. {\updated None of the vulnerabilities detected by CAT were discovered by previous work \cite{mirdita2023cure} or libFuzzer.} Even CAT without CP/IFs was able to discover 15 new vulnerabilities. The graph also illustrates the benefit of using coverage information in the fuzzer, with an improvement of about 27\% after 1000 executions. Our evaluation also finds that our contributions are essential to benefit from coverage-guidance when using batches: Coverage-guided fuzzing without CP/IFs performs almost identically to not using coverage-guidance at all, only achieving a 2\% coverage improvement over the no-coverage approach. This is expected as most batches only contain a few well-performing objects. If the batch is rated as a whole, the coverage-guidance is not able to benefit from the few well-performing objects since it also rates the large amount of bad-performing objects as good. 
To isolate the impact of batching, we additionally evaluate CAT in a single-input configuration (\textit{CAT single}) where each execution processes only one object while retaining all other components (mutation, repair, scoring). This configuration performs significantly worse than the batched version, demonstrating that batching is a key contributor to both coverage growth and vulnerability discovery. \\
\indent \textbf{Benefit of per-object coverage.}
In contrast, object-specific coverage-guidance allows CAT to discover more and \nv{\textit{valuable}} branches that the other approaches overlook. Using CP with IFs allowed our coverage guided fuzzer to find six additional vulnerabilities that we were unable to find without object-specific coverage-guidance, even when running the fuzzer for 24 hours. The vulnerabilities required a specific combination of values in a field as they are located in an edge-case branch that can only be reached when all required fields have the correct value and are at the right location in the object. Without coverage feedback, the fuzzer struggles with finding the exact combination of values that can reach the branch of the edge-case and thereby trigger the vulnerability. One example for this kind of vulnerability in Fort is the signer version tag check, which crashes for a field value that starts with 0x32, terminates with the tag 0x02, and is otherwise well-formatted. Only with this specific configuration will the vulnerable branch be triggered, which could not be achieved without coverage-guidance. \nv{The coverage-guidance allows the fuzzer to discover the necessary combination iteratively, meeting the required conditions in separate batches, exploring the new branches and finally triggering the vulnerability.}\\
\indent Additionally to discovering difficult-to-reach branches, object-specific coverage-guidance also finds more branches over time. As visible in the graph, all other approaches that lack our techniques struggle to construct objects that can penetrate more difficult-to-reach sections of the code, leading to minimal increases after the 500s iteration. In contrast, the coverage-guided fuzzer with CP/IFs continuously finds new branches, even until the 1000th iteration (1 million tested objects), utilizing the feedback to construct values that are more likely to discover new branches.

\subsection{Technical Ablation}
\begin{figure}[t!]
    \centering
    \includegraphics[width=1.0\columnwidth]{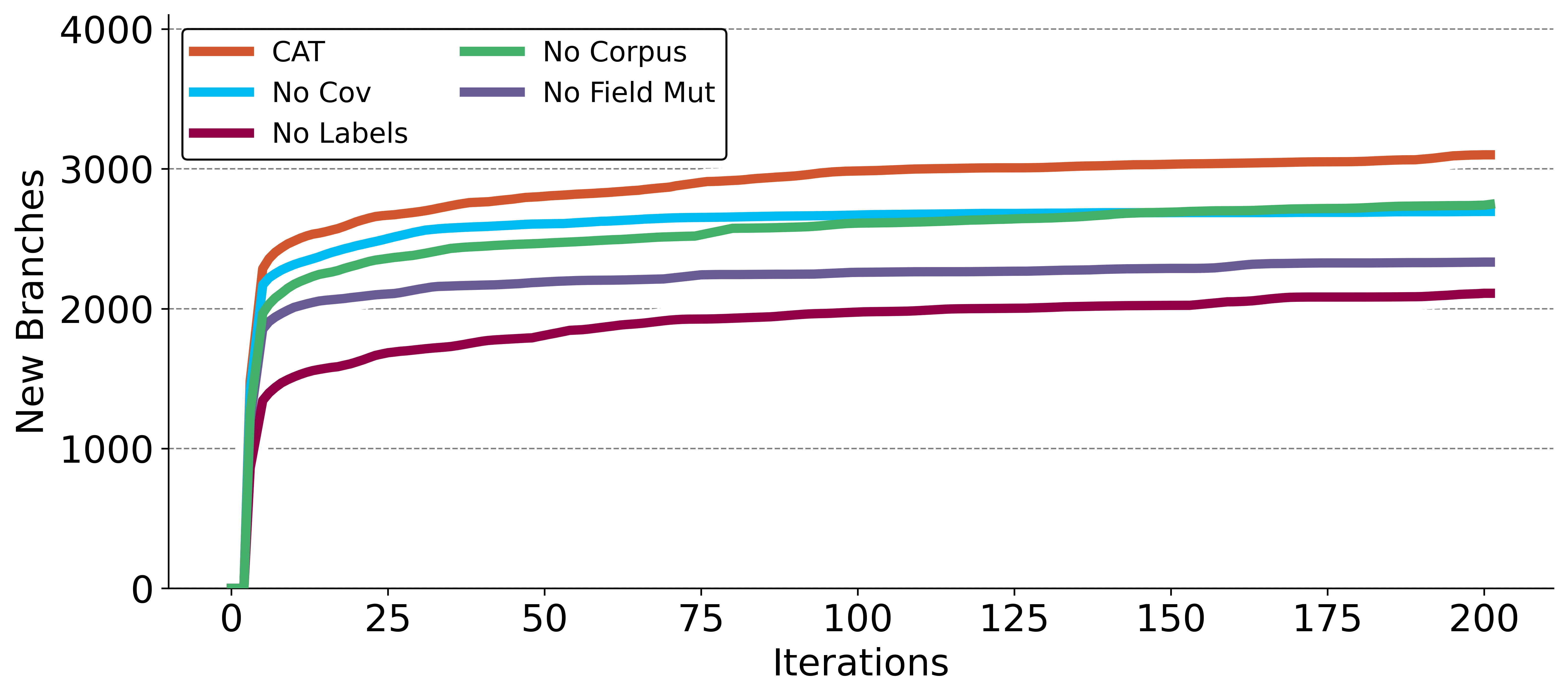}
    \caption{\updated{Ablation coverage increases.}}
    \label{fig:ablation}
\end{figure}

We additionally perform an ablation study to illustrate how each technical contribution adds to the overall performance of CAT, shown in Figure \ref{fig:ablation}. The evaluation shows that the custom ASN.1 parser with labels has the biggest impact on results, as objects fail signature validation and the fuzzer cannot penetrate into validation logic, i.e., the fuzzer mostly tests parsing {\updated when lacking cryptographic repairs.} In this context \textit{no labels} corresponds to disabling the repair mechanism, as labels are required to locate and correctly update dependent cryptographic fields (e.g., signatures, hashes, lengths) after mutation.\\
\indent Over time, the negative impact of no corpus surpasses no coverage, illustrating that coverage allows to still find promising inputs even without a corpus. We additionally observe that the corpus and field mutations both contribute to the final coverage, and removing them significantly reduces performance, with a larger impact of the corpus.\\
\indent In conclusion, object-specific coverage-guidance and custom ASN.1 mutations provide a big advantage for fuzzing the RPs, as it allows finding more branches, and guides the fuzzer towards valuable edge-case branches that contain vulnerabilities impossible to find without a object-specific coverage-guided tool. {\updated Overall, the improvements of \cure\ stem from three complementary factors: (1) batching amortizes expensive repository setup and enables realistic multi-object execution, (2) mutation-and-repair ensures that mutated inputs remain sufficiently valid to reach deep validation logic, and (3) object-specific coverage attribution (CP/IFs) preserves the effectiveness of coverage guidance despite batching. While each component individually improves performance, our results show that the most complex vulnerabilities can only be discovered when all three are combined.}
All technical contributions, including adding a corpus, coverage, labeling, and implementing field-level ASN.1 mutations improve fuzzing performance measurably. {\updated Five vulnerabilities were only detected when combining all contributions.} Without CP with IFs, coverage guidance for large batches only provides a minimal improvement over no coverage at all.

\section{\hspace{2mm}Vulnerabilities in RPKI Validation}\label{taxonomy}
Using \cure\ we detect various problems in the RPs, including 21 unknown vulnerabilities, summarized in Table \ref{tab:vulnerabilities}. All discovered vulnerabilities allow an attacker to manipulate or disable RPKI validation, thereby directly affecting secure and stable routing of the victim AS. {\updated The table shows the range of required iterations over the fuzzer runs required to identify each vulnerability. This provides an indication how difficult, on average, each vulnerability is to detect.}

\subsection{Threat Model}
All discovered vulnerabilities allow direct attacks on RPs. To exploit any of the vulnerabilities, an attacker only needs to setup up a valid RPKI repository. This is feasible for any attacker owning IP resources by registering at their respective RIR. After registration, the RIR repository will delegate any requesting RP to the repository of the attacker, allowing the attacker to target \emph{all} globally running RPs from a single vantage point. For exploitation, the attacker uploads a malicious object into their repository, which is eventually downloaded and processed by all RPs requesting the repository. Since all vulnerabilities enable exploitation of their respective implementation, all have been patched by the developers, except for the deprecated OctoRPKI. We elaborate the impact of the discovered vulnerabilities next.

\subsection{Denial-of-Service}
Denial-of-Service vulnerabilities are a severe issue in RPKI, as the RP is the central component to provide validated RPKI data to BGP routers. If the RP becomes unavailable due to a DoS attack, the routers can not fetch fresh RPKI data and downgrade their protection, falling back to insecure BGP. DoS on RPs thus directly impacts secure routing of all connected routers. This enables a range of follow-up attacks, like eavesdropping traffic, DNS cache poisoning or hijacking web-resources. That DoS is a severe vulnerability in RPKI is illustrated by the CVSS score \textit{high} assigned to all DoS CVEs in this work.
\\
\indent We identified DoS vulnerabilities in all RP clients, resulting from parsing issues, missing bounds checks, and non-graceful failure for validation errors. 
Vulnerabilities persist even in the two clients that use OpenSSL for parsing and validation, Fort and rpki-client. While OpenSSL improves the resilience of parsing and we did not find any vulnerabilities within OpenSSL code, misinterpretation of parsing or validation results still leads to 8 vulnerabilities. For example, rpki-client crashes due to an assert statement following the parsing of the eContentType field. The validation logic assumes that the parsing function should never return a null value. While this is true if parsing of the content itself fails, the parser returns null if the field has a faulty tag. Since rpki-client asserts that the return value must not be null, the execution is terminated for faulty tags. We observe a similar misunderstanding of parsing results in Fort. If an object does not contain any binary data within the eContent field, the parsing function for the eContent returns null. Since Fort does not check for a null return and later tries to access the eContent field within a struct, execution crashes.
These vulnerabilities highlight that while using established libraries for parsing provides a benefit, it does not automatically protect against crashes, and developers need to be aware of possible return values and internal logic of functions they use.

\begin{table}[t!]
\centering
\renewcommand{\arraystretch}{0.7}
\footnotesize
\begin{tabular}{l|c|c|c|c|c}
\textbf{RP}    & \textbf{Vulnerability} & \textbf{Type} & \textbf{Count} & \textbf{Oracle} & \textbf{Iterations}\\ \hline
Routinator &  UTF-8 Cont.  & DoS & 2 & Crash & 300 - 500\\ 
 &   + Name &  &  & \\ \hline
rpki-client    & assert() & DoS & 1 & Crash & \textless 100\\ \hline
rpki-client    & RRDP Stall & DoS & 1 & Time & 100 -- 200\\ \hline
Fort    & Null-Check & DoS & 5 & Crash & 100 -- 500\\ \hline
Fort    & Key Usage & Overfl. & 1 & Crash & 400 -- 800\\ \hline
Fort    & More-Bytes  & DoS & 1 & Crash & 300 -- 500\\ \hline
Fort    & HashList & DoS & 1 & Crash & \textless 100\\ \hline
OctoRPKI & CMS Parsing & DoS & 3 & Crash & 100 -- 400\\ \hline 
OctoRPKI    & Name encode & DoS & 1 & Crash  & 100 -- 200\\ \hline
OctoRPKI & SKI ID & Cache & 1 & Incon & 300 -- 500\\ \hline
Prover & Snap. Item  & DoS & 4 & ROA & 100 -- 600
\end{tabular}
\caption{{\updated Discovered vulnerabilities in RPs.}}
\label{tab:vulnerabilities}
\end{table}

\subsection{Buffer Overflow}
One detected crash in Fort is triggered by a corrupted compiler stack canary getting overwritten due to a buffer overflow. The buffer overflow results from a call to the memcpy function with data and length that are attacker controlled. The vulnerability received a CVSS score of 9.8.\\
\indent To exploit this vulnerability, an attacker creates an RPKI certificate with a custom key-usage bytes and uploads it to an RPKI repository to attack all global vulnerable clients. The length value of this field, which can be set by the attacker, is used as the write length of the memcpy. The attacker can insert arbitrary bytes into the key-usage bytes, and all data after the 12th byte will overflow the allocated buffer and overwrite the value behind the current stack value.

Investigations show that modern built-in compiler and operating system protections, like stack canaries and ASLR, successfully prevent Remote Code Execution (RCE), reducing the bug to a DoS vulnerability. Without protections, we can demonstrate RCE through jumps to libc. Despite limited exploitability when protections are enabled, the developers still decided to register the vulnerability with a critical 9.8 CVSS score. The reason for this precaution is the severe impact \textit{if} exploitation is successful. RPs usually run in a management section of the network, and have direct connections to routers. Thus, besides providing the attacker the ability to manipulate all RPKI data served to the routers, RCE additionally enables a vector to attack the routers directly. Further, relying on protections might fail if an older compiler is used, or if the attacker discovers an additional vulnerability in Fort that, e.g., leaks memory and thus enables copying the canary. Circumventing compiler protections has been demonstrated in previous work \cite{shacham2004effectiveness}.

\subsection{Remote CA Object Poisoning}
We find two new classes of poisoning vulnerabilities; storage attacks that target specific objects, and parsing attacks that invalidate all objects under a given root CA.\\
\indent \textbf{Attacking single object.} We discovered a bug in OctoRPKI that allows an attacker to prevent the processing of one specific object by any other CA. The bug stems from subject name handling in OctoRPKI. The RFC states that these names should be unique for all objects of one CA. Likely following this logic, OctoRPKI enforces \textit{globally} unique subject names for objects. When encountering a subject name OctoRPKI has already seen, the object will not be processed. If the attacker creates an object in their repository with a subject name identical to an object by any other CA, they can achieve a targeted discarding of only the attacked object, while all other objects of the victim CA will validate normally.
The main advantage of this attack is stealth. 
An attacker can disable one specific object which protects the victim IP resources while keeping all other objects intact and the RP running, avoiding detection.\\
\indent \textbf{Attacking root CAs.} A second bug that allows to prevent the processing of a subset of objects from victim CAs is present in Prover. Faulty handling of certain validation failures, like an undefined boolean value or a faulty tag for signatureParameters, leads to the RP discarding all objects under the same root certificate, effectively disabling one sub-tree of the RPKI hierarchy. If the attacker loads a mutated object into its repository registered under RIPE, Prover will not accept any other object by any CA under RIPE. Although this attack is noisier than the single-object attack as more CAs are affected, detection is challenging since the RP merely discards some objects without crashing.\\
\indent \textbf{Discarding valid objects.} We further identify a bug in Fort that leads the RP to treat objects of unknown types differently, depending on whether these objects were downloaded over rsync or RRDP. If rsync was used, objects of an unknown type contained in a repository will lead to Fort discarding the entire repository content. 
We find that discarding repositories with unknown types already leads to problems in real-world RPKI deployments. When running Fort, we found 3 repositories that were downloaded over rsync and had objects of the type ASPA, which is not yet supported by Fort, leading to Fort discarding the repository. \\
\indent \textbf{Long stalling.} Routers rely on timely access to valid RPKI data for their routing decisions. We find that a single repository can stall Fort for a substantial amount of time when creating many child CA certificates, with repository URIs that are not reachable by Fort. The RP will timeout resolution after 4s, and retry resolution once, leading to 8s delay per uploaded child. Since there is no limit on the processing time spent in a branch of the RPKI hierarchy, this long timeout allows an attacker to stall Fort for a substantial amount of time, thereby creating the threat of outdated repository states or worse, expired manifests resulting in discarding of valid objects. 

\section{\hspace{9pt} Future Research}\label{sc:future}
The methods we developed lay the foundation for applying efficient coverage-guided fuzzing to a range of other cryptographic protocols. We define the following criteria to determine whether a given complex fuzzing problem benefits from batching and thereby from our techniques, like IFs / CP:\\
1. Inputs naturally come in repositories or multi-object setups and initialization dominates runtime cost.\\
2.1: Testing inputs as batches yields substantial speed improvements. More precisely, batching provides speedups when each execution incurs a setup or initialization cost that is shared across all inputs in the batch, rather than repeated independently per input. In such cases, expensive operations, such as repository construction, cryptographic validation context setup, or auxiliary object generation, can be amortized across multiple inputs, resulting in sublinear cost growth with respect to the number of inputs.\\
and/or \\
2.2: Batch testing enables more complex setups that are not achievable through sequential fuzzing, thereby increasing coverage.
This applies when the setup involves dependencies or interactions between fuzzing inputs, e.g., when the validation of one object influences the validation of another, and this relationship cannot be captured through sequential single-input runs. \\
3: The setup cannot easily be cached, snapshotted, or forked. This is the case if the setup includes (cryptographic) dependencies on the fuzzing inputs, like hashes or lists of input objects.\\
4: Partial invalidity does not abort processing of the entire batch.
This property is necessary to ensure that a full batch can be processed to completion without validation being aborted prematurely.

These criteria directly reflect the challenges identified in Section \ref{sec:challenges} for RPKI, where high setup overhead, inter-object dependencies and limited caching opportunities make batching particularly effective. Applying these criteria, the techniques we develop for batching with IFs could be used, e.g., to fuzz TCP/IP in different layers simultaneously, or to fuzz large TLS certificate chains. 
Further, we designed our mutation logic to be extendable, allowing future work to use our mutation module to fuzz protocols with ASN.1 structure for objects, like TLS. Our approach of integrating mutation with field adaptions through labeling facilitates fuzzing of other cryptographic protocols, like TLS or DNSSEC \cite{beurdouche2015,chung2017longitudinal}.

\section{\hspace{3mm} Conclusions}\label{sc:conclusions}
Starting as an experimental technology in the late 2000s, RPKI has become a central component in the Internet's defense against routing attacks and already protects a significant share of networks. Today, over {\updated 61\%} of announced prefixes are covered with ROAs \cite{rpkiMonitor}, and about {\updated 30.1\%} of networks enforce ROV \cite{rovsita}. Despite RPKI's critical role in the stability and security of Internet routing, our research discovers 21 previously unknown severe vulnerabilities, which were already assigned 8 CVEs, including high-impact remote code execution exploits and cache-poisoning attacks (CVSS scores of up to 9.8). 

These findings highlight a gap in the security of existing RPKI implementations, demonstrating that even well-established crypto-based defenses are susceptible to severe exploitation. We show that contrary to previous assumptions \cite{mirdita2023cure}, using established parsing libraries like OpenSSL can lead to errors if return values are not handled correctly. Our discoveries emphasize the need for advanced, comprehensive testing to ensure the reliability of critical Internet security protocols.

One key lesson learned is the benefit of overcoming the hurdles to apply coverage-guided fuzzing to complex cryptographic protocols like RPKI, outperforming traditional methods by utilizing execution path feedback to steer the fuzzing process. This approach provides a structured and efficient method for discovering vulnerabilities that would otherwise go unnoticed, paving the way for more robust and secure implementations of critical Internet infrastructure.

Furthermore, the methodologies we developed are not limited to RPKI. The principles of our coverage-guided, non-sequential fuzzing approach, as well as our ASN.1 object mutation techniques, are broadly applicable and can be adapted to improve the security of other cryptography-based protocols, such as TLS and DNSSEC. We open-source our tools and techniques for reproducibility and to enable the research and operational communities to build on them in future work. 

\section*{Acknowledgments}
{\updated This work has been co-funded by the German Federal Ministry of Education and Research and the Hessen State Ministry for Higher Education, Research and Arts within their joint support of the National Research Center for Applied Cybersecurity ATHENE and by the Deutsche Forschungsgemeinschaft (DFG, German Research Foundation) SFB~1119.}

\section{Ethical Considerations}
We ensure that our research is ethical by following best practices for research in network and software security \cite{bishop2008ethical,dittrich2012menlo,sweeney2015sharing}. All testing and fuzzing activities are performed in controlled environments, isolated from production systems. We responsibly disclose all discovered vulnerabilities to affected vendors and the vulnerabilities were consequently fixed.

{
\footnotesize
\bibliographystyle{IEEEtran}
\bibliography{main,ref,bib,fuzz}
}

\appendices
\section{Meta-Review}

The following meta-review was prepared by the program committee for the 2026
IEEE Symposium on Security and Privacy (S\&P) as part of the review process as
detailed in the call for papers.

\subsection{Summary}
The paper presents an automated approach, embodied in a tool called CAT, for fuzzing Resource Public Key Infrastructure (RPKI) validator implementations. CAT is a greybox, coverage-guided fuzzer designed to support highly structured and cryptographically signed input formats. Compared to prior approaches for fuzzing RPKI validator implementations, CAT introduces input batching as well as fine-grained mutation and repair mechanisms. Together with object-level code-coverage attribution, these features enable CAT to explore deeper parts of the target code and uncover both crash and logic bugs. In its empirical evaluation, CAT uncovered previously unknown vulnerabilities in RPKI validator implementations.

\subsection{Scientific Contributions}
\begin{itemize}
\item 3. Creates a New Tool to Enable Future Science.
\item 5. Identifies an Impactful Vulnerability.
\item 6. Provides a Valuable Step Forward in an Established Field.
\item 7. Establishes a New Research Direction.
\end{itemize}

\subsection{Reasons for Acceptance}
\begin{enumerate}
\item CAT features a fine-grained, structure- and encoding-aware mutation engine for inputs encoded as ASN.1 values. This is a technically useful capability that may prove valuable beyond RPKI, especially for fuzzing targets whose inputs are ASN.1-encoded.
\item CAT supports batching multiple RPKI objects into a single test case while attributing code coverage to individual inputs. This design appears to improve fuzzing throughput and exploration depth relative to prior approaches that process one input at a time.
\item CAT uncovered 21 previously unknown vulnerabilities in existing RPKI validator implementations, including implementations that had already been tested by prior work.
\end{enumerate}

\subsection{Noteworthy Concerns} 
None.

\end{document}